\documentclass[journal=jacsat,manuscript=article]{achemso}

\usepackage[version=3]{mhchem} 

\usepackage{xcolor} 
\usepackage{caption}
\usepackage{subcaption}
\usepackage{amssymb}
\usepackage{ulem}

\author{Xhorxhina Shaulli}
\affiliation{Department of Physics, University of Fribourg, Chemin du Musée 3, 1700, Fribourg, Switzerland}
\altaffiliation{Equal first author}
\author{Rodrigo Rivas-Barbosa}
\affiliation{Department of Physics, Sapienza University of Rome, Piazzale Aldo Moro 2, 00185 Roma, Italy}
\altaffiliation{Equal first author}
\author{Maxime Jolisse Bergman}
\affiliation{Department of Physics, University of Fribourg, Chemin du Musée 3, 1700, Fribourg, Switzerland}
\author{Chi Zhang}
\affiliation{Department of Physics, University of Fribourg, Chemin du Musée 3, 1700, Fribourg, Switzerland}
\author{Nicoletta Gnan}
\affiliation{CNR Institute of Complex Systems, Uos Sapienza, Piazzale Aldo Moro 2, 00185, Roma, Italy}
\alsoaffiliation{Department of Physics, Sapienza University of Rome, Piazzale Aldo Moro 2, 00185 Roma, Italy}
\author{Frank Scheffold}
\email{frank.scheffold@unifr.ch}
\affiliation{Department of Physics, University of Fribourg, Chemin du Musée 3, 1700, Fribourg, Switzerland}\altaffiliation{Equal senior author}
\author{Emanuela Zaccarelli}
\email{emanuela.zaccarelli@cnr.it}
\affiliation{CNR Institute of Complex Systems, Uos Sapienza, Piazzale Aldo Moro 2, 00185, Roma, Italy}
\alsoaffiliation{Department of Physics, Sapienza University of Rome, Piazzale Aldo Moro 2, 00185 Roma, Italy}
\altaffiliation{Equal senior author}

\title[\textsf{achemso} demo]
{Probing temperature-responsivity of microgels and its interplay with a solid surface  
by superresolution microscopy and numerical simulations}

\abbreviations{IR,NMR,UV}

\begin{document}

\begin{tocentry}

\begin{center}
   \includegraphics[width=7.5cm]{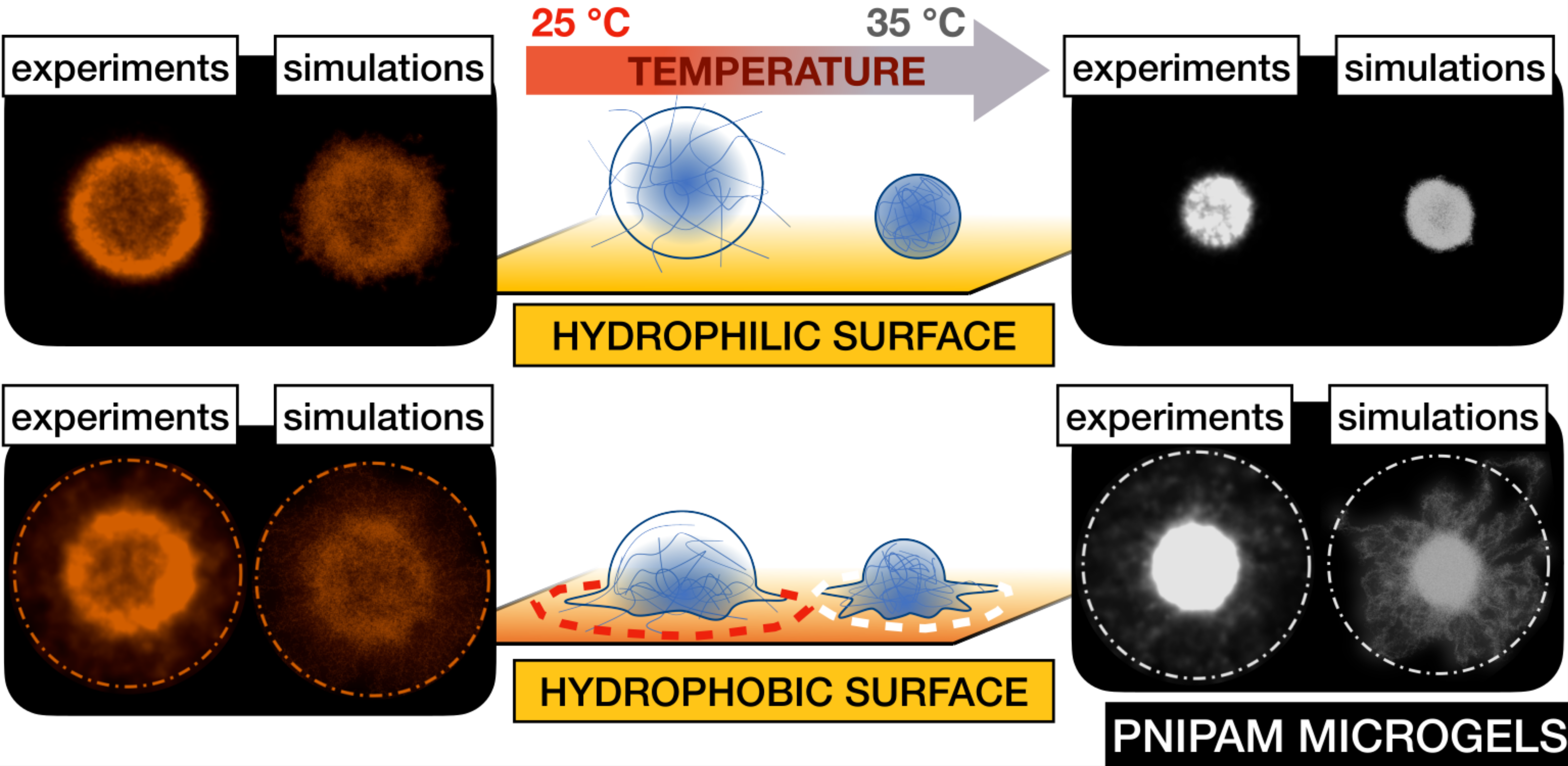}
\end{center}





\end{tocentry}

\begin{abstract}
Superresolution microscopy has become a powerful tool to investigate the internal structure of complex colloidal and polymeric systems, such as microgels, at the nanometer scale.
The ability to monitor microgels response to temperature changes {\it in situ} opens new and exciting opportunities to design and precisely control their behaviour for various applications. When performing advanced microscopy experiments, interactions between the particle and the environment can be important. Often microgels are deposited on a substrate since they have to remain still for several minutes during the experiment. This study uses dSTORM microscopy and advanced coarse-grained molecular dynamics simulations to investigate, for the first time, how individual microgels anchored on hydrophilic and hydrophobic surfaces undergo their volume phase transition in temperature. We find that, in the presence of a hydrophilic substrate, the structure of the microgel is unperturbed and the resulting density profiles quantitatively agree with simulations performed in bulk conditions. Instead, when a hydrophobic surface is used, the microgel spreads at the interface and an interesting competition between the two hydrophobic strengths --- monomer-monomer vs monomer-surface --- comes into play at high temperatures. The remarkable agreement between experiments and simulations makes the present study a fundamental step to establish this high-resolution monitoring technique as a platform for investigating more complex systems, being these either macromolecules with peculiar internal structure or nanocomplexes where molecules of interest can be encapsulated in the microgel network and controllably released with temperature.
\end{abstract}

\section{Introduction}
One of the fascinating aspects of colloidal science is the possibility of investigating mesoscopic particles with experimental techniques that are able to resolve their structure and dynamics at the single particle level. 
Thanks to real-space particle tracking, colloidal particles with size ranging from hundreds of nanometers up to a few microns have long been used as suitable benchmarks to test theories and numerical models. This approach has been successful for a long time, bringing important experimental contributions to fundamental problems as, for example, the nucleation and growth of colloidal crystals~\cite{gasser2001real}, the structural relaxation dynamics close to the glass transition~\cite{weeks2000three,hallett2018local} or the 2D melting scenario of hard colloids~\cite{thorneywork2017two}. Recently, however, the rapidly-growing field of ``smart'' materials, i.e. materials designed to respond to external stimuli, has moved the community's attention from standard ``hard-sphere'' colloids to softer particles, mainly of polymeric nature and with a complex internal architecture. Within this category  microgels are among the most studied colloidal systems. Microgels are cross-linked polymer networks whose properties are intimately related to the type of polymer they are made of and to the topology of the network. Their complex internal architecture provides  particles with an internal elasticity that allows them to shrink, deform and interpenetrate to some extent. Even more intriguing is their ability to respond to external stimuli: depending on the type of polymer employed in the synthesis, microgels can adjust their size in response to temperature, light, or pH changes, to name a few~\cite{pelton2000temperature,Stieger2004, schimka2017photosensitive}.
A well-known example is that of poly- (N-isopropylacrylamide) (pNIPAM) based microgels that exhibit conformational changes in response to solvent composition or temperature. Specifically, they undergo a reversible volume phase transition (VPT) at a temperature of about 32~\textdegree C that leads to network shrinking and, consequently, to a change in the size of the particle~\cite{wu1998globule,saunders1999microgel,senff1999temperature,karg2009new}.
Additionally, it is worth stressing that synthesis advances allow tuning of the microgels size and architecture, as well as decoration and functionalization, all of which in the end influence the responsiveness to external cues~\cite{Agrawal2018synthesis,saunders1999microgel} and the mutual effective interactions between the particles~\cite{rivas2022pnipam-peg}.\\
All these features provide microgels a richer behavior compared to standard ``hard-sphere'' colloids: in particular, the possibility to tune their properties in a controllable way makes them suitable as building blocks for designing materials that can be used for several purposes, from understanding fundamental problems in physics~\cite{saunders1999microgel, alsayed2005premelting, zhang2009thermal, yunker2014physics, philippe2018glass} to industrial applications as, for example, viscosity modifiers, tunable optical scattering components or carrier agents~\cite{das2006microgels,fernandez2011microgel,zeng2021programmable}.
In this context, understanding how the internal complexity of microgels at the nanoscale translates into specific macroscopic material properties is one of the main goals of soft matter physics. This requires the use of novel experimental techniques that go beyond standard optical microscopy and push the resolution below the microscale.\\
To this aim, the recent scientific and technological breakthrough in super-resolved fluorescence microscopy (SRFM) \cite{Heilemann2008dSTORM,huang2009super,betzig2014nobel} has had a direct impact on many research areas in biology and materials science. The advanced techniques were soon embraced as valuable characterization tools with unprecedented accuracy and specificity using multi-color fluorescent labeling protocols~\cite{henriques2011palm,buckers2011simultaneous,conley2017jamming}. Compared to biological studies, applications of SRFM progressed slower in investigating colloidal and polymer systems. More recently, however,  it has been shown to be a unique and valuable imaging and characterization method~\cite{busko2012new,Conley2016,woll2017super,gelissen20163d,pujals2019super,scheffold2020pathways}. In contrast to other characterization methods, such as  X-ray scattering or atomic force microscopy (AFM), SRFM provides single-particle information on the nanoscale and chemical specificity at the same time.\\
In the past years, SRFM combined with other methods, like computer simulations and light scattering techniques, have been used to tackle many unanswered questions regarding the microgel network and its properties. In particular, Conley et al. demonstrated the successful application of direct Stochastic Optical Reconstruction Microscopy (dSTORM) to investigate stimuli-responsive pNIPAM microgels~\cite{Conley2016} at different solvent compositions. The same technique was then used in pure water by the same authors to probe the behavior of concentrated microgels suspensions in swollen conditions~\cite{conley2017jamming}. In addition, Bergmann et al. studied microgels with different cross-linking densities using dSTORM via a non-specific labeling approach~\cite{bergmann2018super,otto2020resolving}.  
Different kinds of microgels were studied:  first  , W\"oll and co-workers investigated core-shell microgel particles combining in situ electron and super-resolution microscopy to unravel new levels of structural details of their particles \cite{gelissen20163d}, then they studied pNIPMAM microgels on solid-liquid interfaces below the VPT showing the dependency of the spread of the microgel with the surface degree of hydrophobicity~\cite{Alvarez2019solid} and with the procedure in which the microgel is placed on it~\cite{Alvarez2021deposition}.
None of these previous super-resolution studies have tackled the problem of probing the thermoresponsivity of the microgels directly by changing temperature, in order to directly visualize the occurrence of the VPT in the internal structure of the microgels. The only notable exception is the application of the PAINT technique to core-shell pNIPAM-pNIPMAM microgels~\cite{Purohit2019}, which however only gained information on the radial distribution of the polarity within the particles, rather than the true polymer density profile.\\
Here we fill this gap by providing a dSTORM investigation of the volume phase transition of pNIPAM microgels, 
validating the experimental approach through the comparison with state-of-the-art computer simulations. Indeed, the application of dSTORM technique currently requires the anchoring of the microgels at a nearby surface, that, in principle, could affect the results. The present work shows that the use of a hydrophilic surface allows us to probe the VPT without detectable external perturbation of the microgel structure, both in the swollen and in the collapsed state. This is achieved by a careful comparison with numerical simulations both in the presence and in the absence of the nearby surface.  Next, we actually exploit the presence of the surface to also characterize the VPT of a microgel close to a hydrophobic wall, thus revealing a subtle interplay between surface adhesion energy and hydrophobic interactions within the microgel, whose microscopic mechanism is unveiled by the numerical simulations. Our work thus represents a crucial step in the nanoscopic characterization of thermoresponsive soft particles, paving the way for its use in a variety of different systems.

\section{Results}

\subsection{dSTORM imaging of the Volume Phase Transition of microgels} 
\begin{figure}[ht]
\centering
\includegraphics[width=6.4 in]{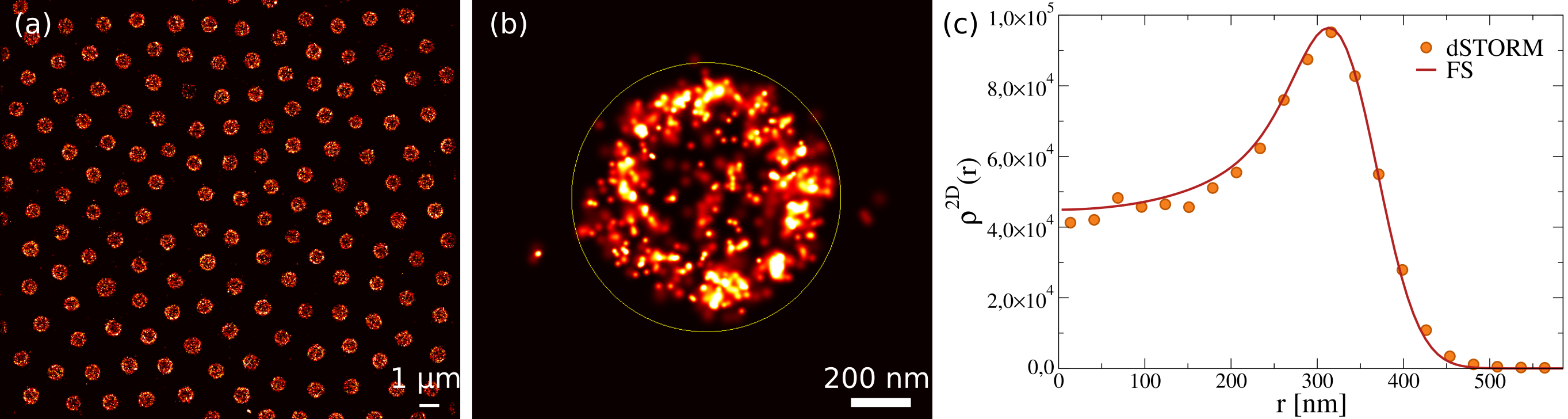}
\caption{(a) dSTORM image of 1.5 mol\% crosslinked, shell labeled, swollen microgel particles at 25\textdegree C . (b) Zoomed individual particle of the same sample. Rendering mode set to individual localization precision, iso (\textit{Picasso}). (c) dSTORM analysis of the microgel radial density profiles illustrating a measured 2D profile (symbols) (averaged over $\sim$ 100 different microgels) and the corresponding fit with the fuzzy sphere model (line).}  
\label{fgr:example2col}
\end{figure}

We investigate pNIPAM microgels with $\sim$1.5 mol\% BIS using dSTORM across the volume phase transition (VPT). Particles are fluorescently labeled in the outer region to facilitate discrimination of different parts of the microgel, namely the core region and the shell including the dangling ends, i.e. the microgel corona. 
Details on the labeling protocol can be found in the Methods Section.
In the following, we denote as {\it shell} the part of the microgel network that is dye-labeled, even if the transition from the dense core to the loosely crosslinked corona is rather gradual and there is no sharp border~\cite{senff1999temperature,scheffold2020pathways}.\\ 
To perform dSTORM experiments, microgels must be anchored to a surface, upon which they are deposited irreversibly by drying and re-suspending them in water~\cite{Conley2016}. In order to appropriately characterize the well-studied phenomenon of the VPT in pNIPAM microgels, we must thus ensure that this anchoring surface does not appreciably interfere with the microgel swelling and deswelling. To this aim, we use a hydrophilic surface, where the coverslip is treated with KOH 3M, sonicated for 10 min, and then exposed for 10 min on a UV-Ozone cleaner, obtaining surfaces with contact angle less than 20\textdegree, as shown in Fig.~S1 of the Supplementary Information (SI). For the imaging process we induce stochastic blinking of the fluorophores as described in the Methods Section. For each dSTORM image, we acquire 30000 - 60000 frames with an exposure time of 10 - 20 ms and proceed with image analysis using the open-source \textit{Picasso} software~\cite{Schnitzbauer2017}, as detailed in SI, following Conley and coworkers~\cite{Conley2016,conley2017superresolution}. Under these conditions we obtain an experimental resolution of around 30 nm in the plane (Figs.~S2-S5).  In Fig.~\ref{fgr:example2col} we show the typical experimental workflow for all microgel particles. We extract quantitative information from dSTORM using a customised MATLAB (MathWorks, USA) routine to determine the 2D fluorophore density profiles $\rho^{2D}(r)$ of the microgels as previously explained  in Ref.~\citenum{Conley2016}. In order to  describe the radially decaying density of the microgel under swollen conditions, the classic fuzzy sphere model is used\cite{Stieger2004}, adapted to the case where microgels solely contain fluorophores in their outer shell, with their denser core being `invisible' in dSTORM experiments, as discussed in the SI (Fig.~S6). The fuzzy-sphere model assumes that a microgel is made of a dense inner core of size $R$ and a fuzzy shell of thickness $2\sigma_{surf}$. Both parameters can be obtained from a fit of the 2D density profiles taken from dSTORM thus providing an estimate of the radius of the particle, $R_{tot}=R+2\sigma_{surf}$, projected on a plane. An example of the corresponding fit of the 2D density profile is reported in Fig.~\ref{fgr:example2col}(c), showing good agreement with the dSTORM data.

\begin{figure}[ht]
  \includegraphics[width=4.8 in]{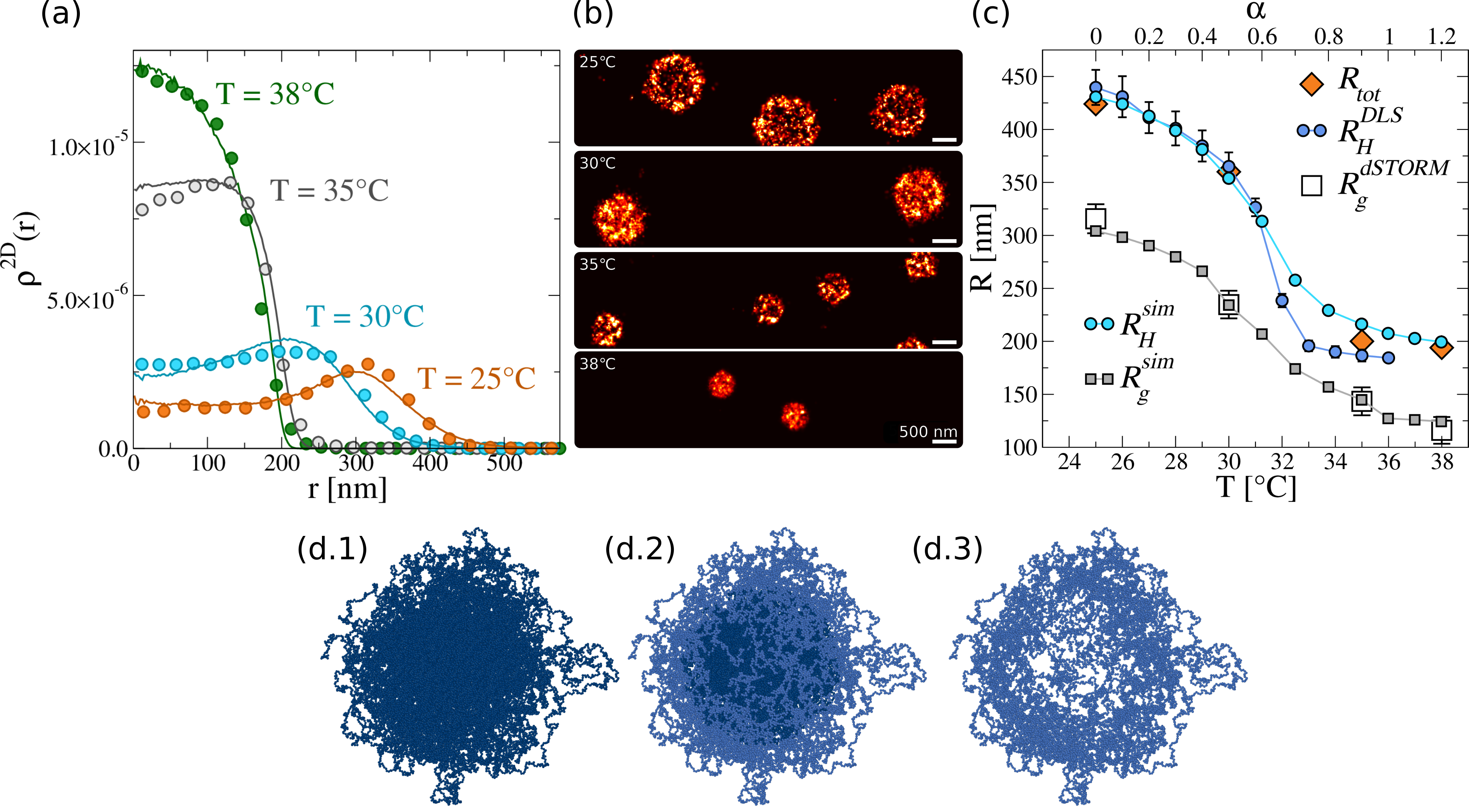}
  \caption{(a) Experimental 2D density profiles (symbols) for microgels at 25, 30 35 and 38~\textdegree C and numerical 2D density profiles (solid lines) calculated with $\alpha_{\text{mm}}=0.0, 0.5, 0.9$ and 1.2, which best match the experimental profiles at the four studied temperatures. In particular, data at $T=25$~\textdegree C  are described without any noise on the fluorophore location ($\sigma_{\text{sd}}=0.0$), while for $T=30$~\textdegree C  $\sigma_{\text{sd}}=0.3$. At $T=38$~\textdegree C, the microgels show a full collapse characterized by the complete absence of a hollow structure and are thus compared to the whole simulated microgel. Finally, for $T=35$~\textdegree C we adopt a mixture of hollow ($\sigma_{\text{sd}}=0.4$) and full microgels, as described in the text and SI (Fig.~S9) for further details. In order to compare experimental and simulation density profiles, we normalize their area integral to one. The simulated curves are reported on experimental units matching the respective 2D gyration radii, as explained in the text; (b) dSTORM images of individual microgels at 25, 30, 35 and 38~\textdegree C (top to bottom). For the complete captured region of interest (ROI), see Fig.~S3;  (c) Comparison of dSTORM estimated radius $R_{tot}$ as a function of temperature with the DLS  $R_H^{DLS}$  and numerical  $R_H^{sim}$ hydrodynamic radii. Also, 2D gyration radii from dSTORM  $R_g^{dSTORM}$ and simulations $R_g^{sim}$ are reported;  (d) Monomers-to-fluorophores conversion process: (d.1) Initial full microgel, (d.2) monomers distinction according to their position either inside or outside the core-shell interface and (d.3) final converted fluorophores. }
  \label{fgr:radialdensityprofile} 
\end{figure}

The novel implementation of a temperature controller in the super-resolution setup allows us to observe the VPT of microgels {\it in situ}. We perform experiments at four different temperatures 25, 30, 35 and 38~\textdegree C, covering the range where the transition occurs. The corresponding 2D density profiles are reported in Fig.~\ref{fgr:radialdensityprofile}(a), showing the typical deswelling behavior of the microgels across the VPT, also visible in the images of Fig.~\ref{fgr:radialdensityprofile}(b). From the fuzzy sphere fits, we find that the microgel 2D radius $R_{tot}$ (see fig.~S6) decreases from $R_{tot}\sim 424$ nm to $R_{tot}\sim 194$ nm within the investigated temperature range. Moreover, the size of the microgel estimated by dSTORM is found to be in good agreement with the hydrodynamic radius $R_H^{DLS}$ measured by dynamic light scattering (DLS), as reported in Fig.~\ref{fgr:radialdensityprofile}(c).\\ 
To understand whether the anchoring to the hydrophilic surface may affect the swelling behavior of the microgels, we compare the experimental data with simulations of a realistic model of microgels~\cite{gnan2017silico},  having the same nominal crosslinker concentration of the experimental sample.  

We start for simplicity by showing results for simulations performed in bulk conditions, i.e. in the absence of a nearby surface. To appropriately model the deswelling of the microgel, we use an effective solvophobic potential, which controls the monomer-monomer attraction through an effective temperature $\alpha_{\text {mm}}$, as done in previous works~\cite{ninarello2019modeling} and described in the Methods Section.  In order to obtain a meaningful comparison between experiments and simulations we mimic \textit{in silico} the distribution of fluorophores  detected by dSTORM. This is done by defining a smooth core-shell interface and converting to fluorophores all monomers belonging to the shell. The procedure ensures that the fluorophores are smoothly distributed across the interface as in the experiments, giving rise to a hollow profile as shown in Fig.~\ref{fgr:radialdensityprofile}(d). More details on the protocol employed to select the fluorophores can be found in the Methods Section and in the SI. We then calculate, as in experiments, the 2D radial density profiles of the simulated microgel across the VPT, only taking into account the external monomers and averaging over all three directions in order to improve statistics. 

The resulting $\rho^{2D}(r)$ are reported in Fig.~\ref{fgr:radialdensityprofile}(a), showing a nearly quantitative agreement with experiments for all the main features: the mass of the core, the extension of the corona as well as the presence of a peak (and its position) at low temperatures. Such a peak is a feature stemming from the combination of the 2D nature of the data and the fluorophore shell labeling, as shown in Fig.~S7. At temperatures below the VPT, we obtain a very good description of the density profiles with the addition of a small noise, that can be attributed to the finite optical resolution of dSTORM (see also Methods Section and SI, Figs.~S8-S9). This is further illustrated in the fluorophore distributions with respect to their distance from the center of mass of the microgel, reported in Fig.~S10, which show that, as $T$ increases, the employed fluorophore profiles move more and more toward the center of mass of the microgel. 
The situation becomes more complex at higher temperatures. First of all, we find that at $T=38$~\textdegree C, the experimental profiles coincide with those of the full simulated microgel. Indeed, under these conditions the microgels are fully collapsed and there is no longer evidence of a hollow structure from the images of Fig.~\ref{fgr:radialdensityprofile}(b).
We interpret the latter as a consequence of several possible effects due to the twofold decrease in particle size above the VPT. First, the size of the microgel's central region, where we expect a depleted fluorescence signal, is significantly reduced and shifted to a small area around the center of the microgel 2D projection. For small $r$-values, however, the radially averaged dSTORM signal is very noisy due to a small number of fluorophore localisations.
Moreover, the depth of field and the microgel size are comparable, which may lead to subtle changes in imaging conditions upon shrinkage~\cite{Conley2019}. In addition, we should consider the possibility that shrinkage leads to a slight redistribution of the dye-labeled polymer strands across the volume, blurring the hole signal. Furthermore, the refractive index of the microgels increases to a degree where we cannot entirely neglect the scattering and attenuation of the exciting light beam. Deciphering the various contributions to the disappearance of the apparent hole is beyond the scope of this paper but will be addressed in future work.
\newline Interestingly, at $T=35$~\textdegree C, just above the VPT, we find the presence in the samples of fluctuations within the different microgel reconstructions, so that some of them clearly display a hole, while some others do not, which could be due to the fact that the experimental situation, e.g. the illumination or the inclination, varies somewhat, as well as the large experimental uncertainty near $r=0$. We consider this experimental observation in our modeling as discussed in detail in the SI (see Fig.~S9), which leads us to separately account for the profiles of the two microgel populations and appropriately mix them in the right proportion in order to obtain the 35~\textdegree C profile shown in Fig.~\ref{fgr:radialdensityprofile}(a).

Remarkably, it is evident from Fig.~\ref{fgr:radialdensityprofile}(a) that the decay at large distances is extremely well captured at all temperatures, signaling that the simulations are able to fully grasp the amount of shrinking observed in experiments. To this aim, we note that 
a previous work~\cite{ninarello2019modeling} had established a mapping between temperature and solvophobic parameter $\alpha_{\text{mm}}$ by comparing numerical form factors with experimental ones of PNIPAM microgels obtained by small angle X-ray scattering. Using dSTORM, we now confirm similar values of $\alpha_{\text{mm}}$ in the studied temperature range. Small deviations occur probably due to the differences in the synthesis process. To convert numerical into experimental units, we  impose that the 2D radius of gyration of the microgel at $\alpha_{\text{mm}}=0.0$, i.e. in the absence of any monomer-monomer attraction, has to be equal to the experimental value obtained at the lowest temperature $T=25$~\textdegree C~\cite{footnote}. With this procedure, we find that the numerical unit length $\sigma$, corresponding to the size of a monomer bead in the model, corresponds to $\sim 9\text{ nm}$ in real units. Such a conversion is then used at all temperatures and throughout the manuscript. The resulting numerical and experimental 2D gyration radii, $R_g^{sim}$ and $R_g^{dSTORM}$ respectively, are found to be in good agreement at all temperatures, as reported in Fig.~\ref{fgr:radialdensityprofile}(c). In addition, we also calculate the hydrodynamic radius $R_H^{sim}$ numerically using the Zeno algorithm~\cite{chremos2022influence}, which was recently  validated for microgels~\cite{del2021two, elancheliyan2022role}, and again we find an overall satisfactory agreement between experiments and simulations. The slight discrepancy observed at high temperatures with respect to the DLS data may likely be due to the fact that the dSTORM buffer solution is different from the one used in DLS measurements. Indeed, for the same buffer conditions as in dSTORM, DLS measurements of $R_H$ are not possible at high temperatures, because of the onset of microgel aggregation. 

Overall, the agreement of numerical simulations in the absence of a surface with dSTORM data suggests that the anchoring of microgels in experiments has a negligible effect on the particles structure throughout the volume phase transition. To confirm that this is the case, we also perform simulations in the presence of a nearby hydrophilic surface. Since the interactions between monomers and wall is mainly repulsive, the surface does not affect the results at any temperature, because the microgel always remain relatively far from the surface. We then mimic the experimental procedure by anchoring a small fraction of microgel monomers on the surface, as detailed in the Methods Section. 
By adding the surface, we first need to quantify the effect of the number of anchoring sites, which we have assessed in Fig.~S11, by a direct comparison to experiments. From this, we get a rough estimate of the fraction of monomers bonded to the surface that is found to be smaller than 0.1\%. Using such a value, we find that the wall-anchored density profiles are equivalent to those in bulk as shown in Fig.~S12, confirming that the microgels structure across the VPT remains unperturbed when anchored to a hydrophilic surface. 
Altogether, these results validate the adopted experimental procedure, indicating that they can clearly detect the changes in the internal structure of the particles at different temperatures, paving the way for its application to different systems. In particular, we expect that dSTORM will be very useful to study copolymer microgels with more complex internal architecture where different temperature behaviors of the forming polymers are at play~\cite{hertle2013thermoresponsive,keerl2009temperature,rivas2022pnipam-peg}.

\subsection{Changing the surface affinity: the VPT of microgels close to a hydrophobic surface} 

\begin{figure}[ht]
\centering
  \includegraphics[width=6.4 in]{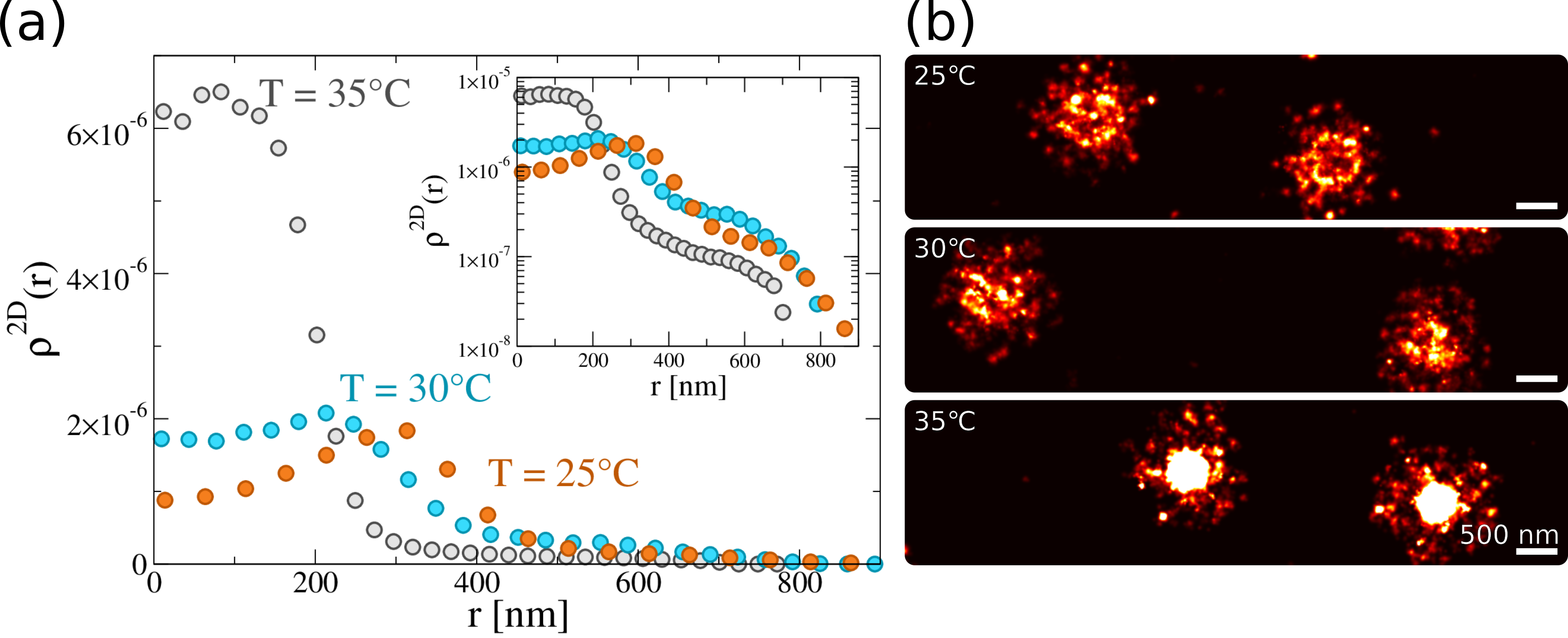}
          \caption{(a) Experimental 2D density profiles for microgels at 25, 30 and 35~\textdegree C on a surface exposed to HMDS. Inset: same density profiles with the y-axis in logarithmic scale to better visualize the tail of the profiles; (b) dSTORM images of individual microgels at 25, 30 and 35~\textdegree C (top to bottom). Here the brightness for the image at 35~\textdegree C is increased to make the anchoring parts visible for the reader. For the complete captured ROI, see SI Fig.~S4.}
   \label{fig:dp2d-phobic-snap} 
\end{figure}

We now discuss the case of microgels close to a hydrophobic surface. To realize this situation, the coverslips were first cleaned with KOH 3M and then exposed overnight to 0.1~mL Hexamethyldisilazane (HMDS).  Contact angle measurements are shown in SI (Fig.~S1), reporting  contact angles larger than 80\textdegree. 
The resulting 2D density profiles  are reported in Fig.~\ref{fig:dp2d-phobic-snap}(a), still showing a clear deswelling of the particles with temperature, however accompanied by the presence of a long tail at large distances, which persists even in the more collapsed conditions. To visualize this behavior, dSTORM images at three studied temperatures are shown in Fig.~\ref{fig:dp2d-phobic-snap}(b), clearly indicating that the microgels adopt a core-shell-like  arrangement, since the external shell tends to maximize the contact with the surface. This behavior is in agreement with previous super-resolution experiments~\cite{Alvarez2019solid}, which were performed at low temperatures only. The present results put forward novel insights on how the tail is maintained even at high temperatures, denoting a large affinity of the microgel to the surface. It is also interesting to compare these results with observations at liquid-liquid interfaces, where microgels adopt the so-called `` fried egg'' configuration,  confirmed by a large amount of experiments (mainly through atomic force microscopy, after deposition onto a surface)~\cite{geisel2012unraveling} as well as numerical simulations~\cite{camerin2019microgels}. Interestingly, in the case of a liquid-liquid interface, temperature effects on the microgel conformation are not very pronounced~\cite{harrer2019stimuli} because of the dominant role of the interfacial tension, as also recently confirmed by direct investigation through {\it in situ} neutron reflectometry~\cite{bochenek2022situ}. Hence, it is worth to examine in more detail the role played by temperature in the present work, where it seems that a competition between hydrophobic interactions, monomer-monomer vs monomer-surface, is at work.

To this aim, we resort to numerical simulations and we model the interactions between monomer and surface with the same potential shape as for monomer-monomer one. In this case, the parameter $\alpha_{\text{ms}}$  controls the affinity between monomers and wall particles. While in the case of hydrophilic surface, we set $\alpha_{\text{ms}}=0$, we now tune $\alpha_{\text{ms}}>0$, to model the hydrophobicity of the surface, as detailed in the Methods Section. We separately consider in the following both the case of a free microgel next to the attractive surface (unbonded), which spontaneously sticks to it, and that of microgel anchored to the surface (bonded). 
After tuning $\alpha_{\text{ms}}$ as shown in Fig.~S13, we find that the best agreement between experiments and simulations is obtained for $\alpha_{\text{ms}}=0.9$, i.e. a relatively large value of the attraction strength, confirming the rather high hydrophobic character of the HMDS surface.
\begin{figure}[ht]
\centering
  \includegraphics[width=6.4 in]{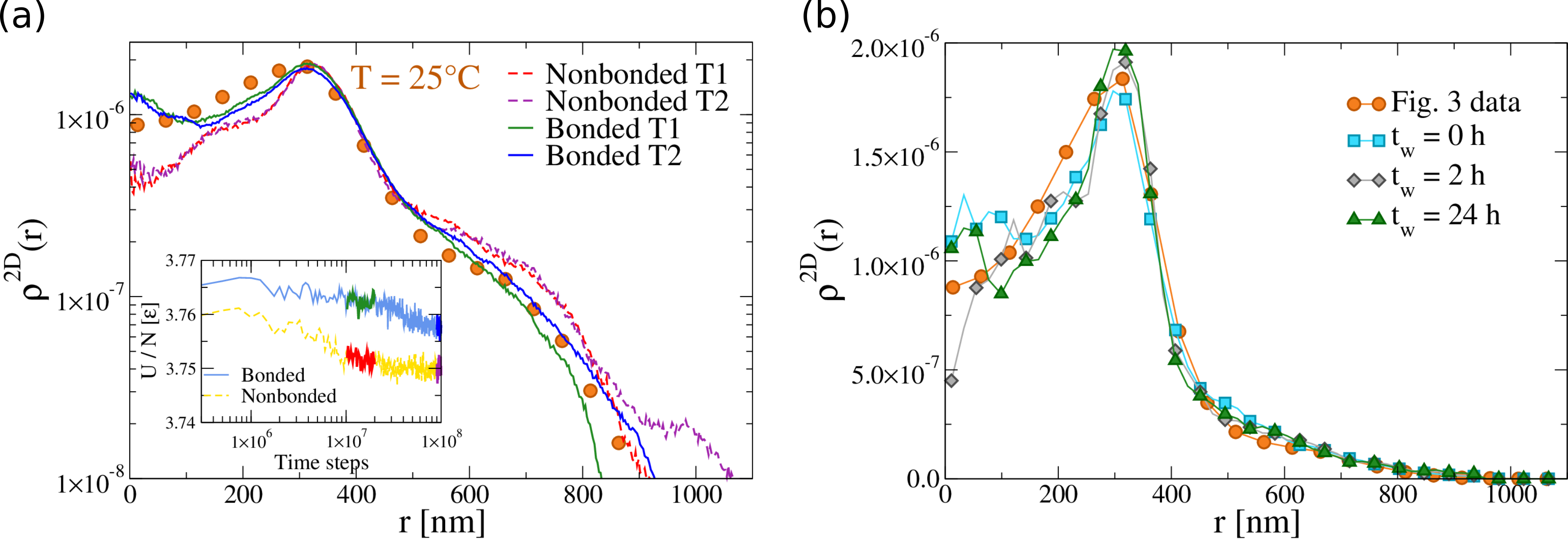}
          \caption{ (a) Experimental 2D density profiles (symbols) for microgels at 25~\textdegree C and corresponding numerical ones obtained with  $\alpha_{\text{mm}}=0.0$ and $\alpha_{\text{ms}}=0.9$ at different times for unbonded (dashed lines) and bonded microgels (solid lines); the time intervals are display in the inset; Inset: Evolution of the potential energy per particle of the bonded and unbonded microgels versus time in numerical simulations. The highlighted regions indicate the time intervals where the correspondingly coloured curves have been calculated in the main panel; (b) additional measurements for 2D density profiles at 25~\textdegree C and waiting times $t_w=$0 h, 2 h and 24 h in comparison to the data reported in Fig.~\protect\ref{fig:dp2d-phobic-snap}(a) and elsewhere in the manuscript. }
   \label{fig:phobic-T25} 
\end{figure}
We compare the resulting numerical and experimental 2D density profiles at $T=$25~\textdegree C in Fig.~\ref{fig:phobic-T25}(a) and note that the system takes a very long time to reach equilibrium. This is illustrated in the inset, reporting the energy per particle versus time, which is found to slowly decrease, denoting a long aging regime, for both bonded and unbonded microgels. During this time, also the density profile of the microgel slowly evolves, accumulating more and more monomers at large distances from the center of mass. This indicates that, for the chosen conditions, the hydrophobic attraction of the monomers to the wall is dominant and pushes the microgel to extend further and further on the surface. Interestingly, from the analysis at different values of $\alpha_{\text{ms}}$, we found that only for $\alpha_{\text{ms}} \gtrsim 0.7$,
the microgel sticks to the surface, otherwise it tends to remain in a spherical (almost unperturbed) condition. However, as soon as the microgel sticks, slow rearrangements of the monomers on the surface take place, giving rise to this non-negligible long-time evolution.
These results show that, on increasing time, the outer shell progressively extends more than in the experiments, thus overestimating the tail of the density profiles in the case of an unbonded microgel. This is also reflected in the short-distance behaviour of the numerical $\rho^{2D}(r)$ whose height results to be lower than the height observed with dSTORM. 
Instead, looking at the numerical data for a bonded microgel, also reported in Fig.~\ref{fig:phobic-T25}(a), we find that the experimental data are much better captured and, in particular, at long simulation times the tail of the distribution roughly extends of the same amount as in experiments. Hence, differently from the case of a hydrophilic surface, here a small degree of anchoring does affect the overall conformation of a microgel due to the strong binding to the surface and the underlying connectivity of the network.
We thus conclude that we cannot neglect the presence of anchoring in the simulations for a more quantitative description of super-resolution data close to a hydrophobic surface.

To verify whether this holds at all $T$, we extend the comparison to higher temperatures to tackle the interesting case of a collapsing microgel close to a hydrophobic surface. To this aim, several details need to be taken into account. The first point to address is aging, which is very pronounced in the simulations, due to the fact that, at high $T$, there is a competition between the monomer-monomer attraction, modelled by $\alpha_{\text {mm}}=0.9$ and the monomer-surface attraction, modelled with $\alpha_{\text{ms}}=0.9$ as determined at $T=25$~\textdegree C. In particular, we find that the monomer-monomer attraction, being augmented by the large number of nearby monomers and the overall connectivity, will eventually dominate when the two $\alpha$ parameters are the same, so that the microgel will tend to collapse further, decreasing the extent of the large distance tail.   
It is now important to assess the role of aging on the experimental results. To this aim, we performed additional measurements for microgels on the hydrophobic surface at different waiting times $t_w$ for $T=25$~\textdegree C. We define $t_w=0$ as the time when we started the dSTORM measurements. When comparing the density profiles of measurements done at $t_w=$0h, 2h and 24h, reported in Fig.~\ref{fig:phobic-T25}(b), we detect no significant differences in the curves. Indeed, the time in between sample preparation and data acquisition, which is roughly of the order of one hour, seems to be long enough for the system to reach equilibrium, so that we can neglect aging effects in the experimentally measured time window. These results should thus be compared with numerical ones at very long times.

\begin{figure}[ht]
     \centering
       \includegraphics[width=3.33 in]{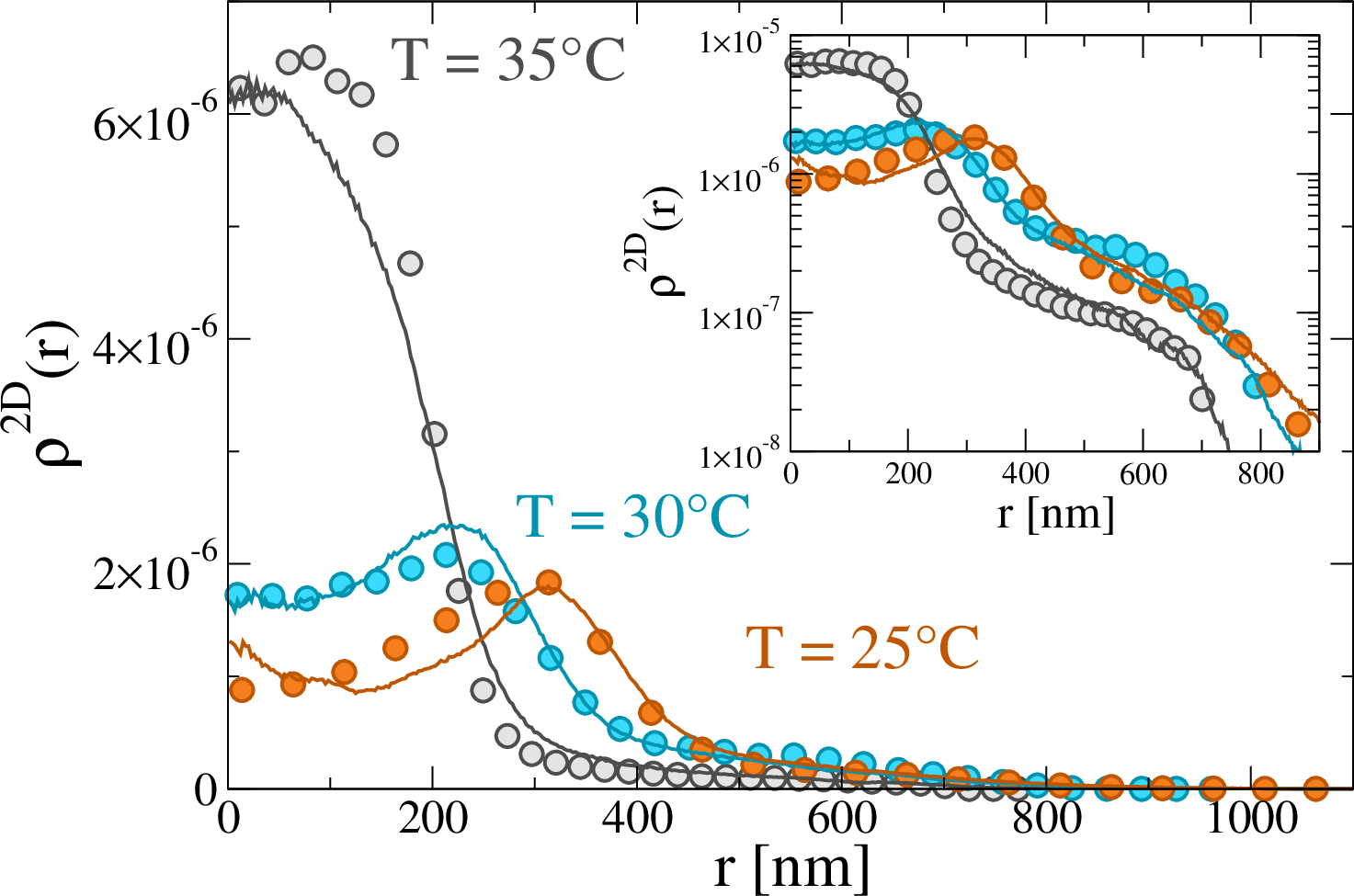}
       	\caption{ (a) Experimental 2D density profiles (symbols) and numerical 2D density profiles (solid lines) calculated with $\alpha_{\text{ms}}=0.9$,  $\alpha_{\text{mm}}=0.0, 0.5, 0.9$. Data for $T=25$~\textdegree C are described without any noise on the fluorophore location  ($\sigma_{\text{sd}}=0.0$), while for $T=30$~\textdegree C  $\sigma_{\text{sd}}=0.3$, in analogy with the hydrophilic case. Instead, data  for $T=35$~\textdegree C show the absence of a hollow structure and are thus compared to the whole simulated microgel. Inset: y-axis in logarithmic scale. }
       \label{fig:phobic-all}
\end{figure}

Next, we consider the fact that, at high $T$  the density profiles may also depend on the way the microgel is anchored on the surface. To address this problem, we consider the anchoring process of a swollen microgel (see Methods) both onto a hydrophilic and onto a hydrophobic surface. This yields different bonding patterns: when the anchoring process is done on a hydrophobic surface the bonds tend to be made for monomers located further away from its plane projected center of mass. We then calculate $\rho^{\text{2D}}(r)$ above the VPT, (see SI, Fig.~S15) and find that the presence of a hydrophobic surface allows for the formation of bonds over a more extended region compared to the hydrophilic one, which results in a larger tail of the density profile. This is in closer agreement with experiments, because it also more realistically mimics the way that the anchoring is made.

Having established the optimal ways of comparing experiments and simulations, we are now able to finally report the comparison of experimental and numerical density profiles at all investigated temperatures close to the hydrophobic surface in Fig.~\ref{fig:phobic-all}. 
Below the VPT temperature ($T$=25 and 30~\textdegree C) the profiles still show the presence of a peak, indicative of the fluorophore outer shell distribution, that we keep exactly identical to that used in the presence of the hydrophilic surface (see Fig.~S10). However, at the highest studied $T$=35~\textdegree C the experimental profile shows a monotonic decrease at low distances, being characterized by the absence of a hollow structure, similarly to what observed at high temperatures for the case of a hydrophilic surface. Indeed, we find that the comparison with any fluorophore distribution is not able to reproduce the experimental data (see Fig.S15), but instead we consider the full microgel profile and obtain a very satisfactory agreement. This happens at a slightly lower temperature with respect to the hydrophilic case, probably due to the presence of the attractive surface, which effectively favours an anticipated collapse of the microgel. 
Overall, Fig.~\ref{fig:phobic-all} shows that simulations, taking appropriately into account the subtleties of anchoring and fluorophore detection, can capture the experimental data at all temperatures also in the case of a hydrophobic surface.

\begin{figure}[ht]
\centering
  \includegraphics[width=6.4 in]{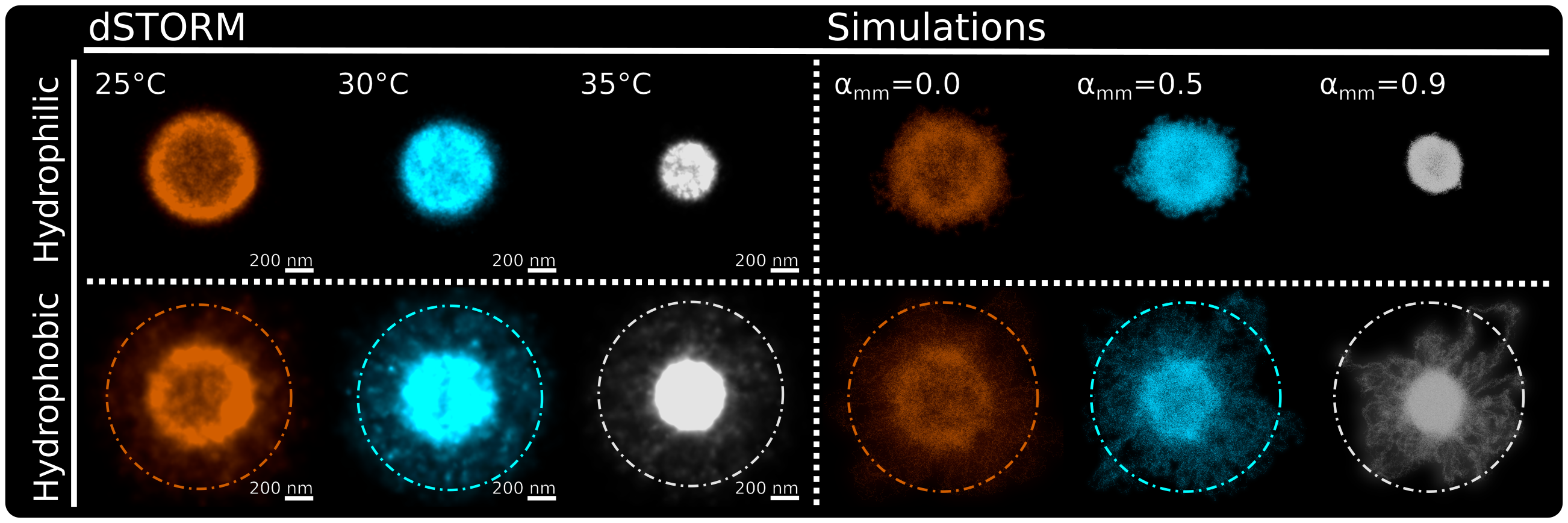}
  \caption{Top panel - averaged images of individual microgels for increasing temperatures (25, 30 and 35 \textdegree C) on hydrophilic surface (dSTORM-left, MD-right).
  Bottom panel - averaged images of individual microgels for increasing temperatures (25, 30 and 35 \textdegree C) on hydrophobic surface (dSTORM-left, MD-right). The brightness for the images on hydrophobic surface is increased to make the anchoring parts clearly visible for the reader.}
 \label{fig:h-vis}
\end{figure}

The quantitative comparison between experiments and simulations can be summarized in Fig.~\ref{fig:h-vis}. This reports the average images, recorded by dSTORM, of individual microgels at three studied temperatures and close to the two different surfaces. While for the hydrophilic case, the typical spherical pattern is retained, for the hydrophobic surface, the shell around the core persists at all temperatures, in very good agreement between experiments and simulations. In addition, it is clear from the direct comparison of the images for  35 \textdegree C for the two examined cases, that the microgel internal structure is much more compact (without a perceptible hole in the centre) for the hydrophobic conditions, in agreement with the employed numerical description.

\section{Conclusions} 
 In this manuscript we performed a detailed investigation of the volume phase transition of standard pNIPAM microgels by dSTORM measurements combined with coarse-grained numerical simulations of realistic microgels. The experiments were carried out by anchoring the microgels to two different substrates: a fully hydrophilic and a largely hydrophobic surface. 

In the former case, we find that the presence of the nearby wall does not significantly affect the conformations of the microgels, which always remain spherical at all investigated temperatures. It was thus possible to monitor the occurrence of the VPT and to quantitatively compare with simulations of isolated microgels in bulk and on hydrophilic surfaces. The comparison was performed adopting a procedure where only a fraction of monomers, those emulating the fluorophores, were taken into account in the calculation at low temperatures. Instead, with the increasing collapse of the microgel, the hollow core progressively becomes less experimentally detectable, so that the profile is well captured by that of the full microgel.
The favourable agreement of experiments and simulations, on one hand, reinforces the numerical model, previously shown to be able to describe the evolution of microgels form factors across the VPT~\cite{ninarello2019modeling} and now validated also in direct space, and on the other hand, establishes  the use of the temperature-controlled dSTORM approach used in this work for the nano-scale resolution of the internal structure of microgels. Indeed, despite the technique requires the anchoring onto a surface, the successful comparison with simulations indicates that such anchoring can be safely neglected in the description, because it does not affect the overall swelling-deswelling process. To this aim we showed that the dSTORM measurements are quantitatively captured also by simulations performed in the absence of a surface.

Having validated the combination of our methods, we then moved on to examine the influence of the surface on the VPT when interactions between monomers and substrate are highly attractive. For the latter case, we are not aware of any previous detailed characterization of the microgel conformation attached to a hydrophobic surface as a function of temperature. While at low temperatures, in agreement with previous works~\cite{Alvarez2019solid,Alvarez2021deposition}, the preferential attraction to the surface induces the microgel to spread and deform, rather similarly to the case of liquid-liquid interfaces~\cite{destribats2011egg,camerin2019microgels}, thus adopting a core-shell-like structure, the previously unexplored high temperature regime reveals an interesting behavior. This is due to the emerging competition between microgel deswelling, controlled by the decrease in the solvent-monomer affinity, and the adhesion to the surface, modulated by the hydrophobic coating of the substrate. Thanks to the help of simulations, where these two parameters can be varied more easily than in experiments, a delicate interplay between the two mechanisms arises, giving rise to a subtle aging behavior in the simulations. However, in experiments, the typical preparation time is about one hour, which allows for a full equilibration of the system, so that we can safely neglect aging effects in the measurements. In the simulations, we instead find that a very long equilibration takes place, which does not change the overall shape of the density profiles but may affect the tail of the profiles. To determine the long-time behavior of such tails, it turns out to be crucial to appropriately consider the presence of the anchoring of the microgels to the substrate. In particular, in the absence of anchoring, our simulated microgels would continue to spread even further at the studied conditions, a process that is inhibited by the presence of permanent bonds with the surface which, combined with the network structure of the particles, limits the growth of the tail at long times, in good qualitative agreement with experiments. Importantly, comparing the present results for the hydrophobic surface with those obtained at liquid-liquid interfaces, we note that the use of a solid substrate enables a more effective exploitation of temperature effects on the final microgel configuration. Indeed, we find a rather exotic conformation made of a collapsed-core plus an extended shell due to the interplay between the two tunable and competing hydrophobic strengths.  Such conformation is not easily observed at liquid-liquid interfaces, because in that case the interfacial tension is always dominant with respect to temperature. 

In future works it will be interesting to extend the present results to the investigation of different surfaces, varying the hydrophobic affinity of the microgel~\cite{scheffold2020pathways} as well as to vary the crosslinker concentration. While here we focused on rather soft microgels with low amount of crosslinkers, we expect that a variation of the particle softness~\cite{scotti2022softness} will produce additional interesting features. In particular, the use of ultra-low-crosslinked microgels may be especially worth exploring, due to their intrinsic difference with respect to standard ones~\cite{bochenek2022situ,schulte2019ultralow}. In addition, a careful comparison between microgel conformations at liquid-solid vs liquid-liquid interface, building on the one performed by AFM measurements~\cite{schulte2019ultralow}, would be desirable to fully unveil the conformational differences of the particles with nanoscopic resolution.

The present study opens the way to use dSTORM to investigate the VPT behavior of microgels of complex inner structure, including co-polymerized ones with different responsivity with respect to temperature or pH or different internal architectures, e.g. interpenetrated network microgels. To this aim, it would be desirable to first extend the present analysis to different labeling over the whole microgel and not only on the surface, as previously done in Ref.~\citenum{Conley2016}, to be able to detect variations throughout the particles, even in the core. In addition, a full 3D imaging should be implemented so that the full density profiles could be directly compared to simulations or to experimental form factors, although this is not crucial in the presence of microgel deformation onto a surface, such as the one studied in this work.

Finally, an additional step foward will be to move from simple microgels to compartmentalized ones~\cite{gelissen20163d} useful for segregating reactive components and  coordinating chemical reactions, or to nanocomplexes where these are decorated with other objects, such as nanoparticles, e.g. for enhancing plasmonic or optical properties\cite{Karg2007}, or biomolecules, e.g. for delivery purposes~\cite{Pergushov2021,Chen2021,dave2022microgels}. In these cases, the advanced superresolution approach developed in the present work will enable the visualization of the full temperature behavior {\it in situ}, which is crucial to  control adsorption and release of these molecules with high potential for their fundamental mechanisms occurring at the nanoscale and for improving their applications in different fields.

\section{Materials and Methodology}

\subsection{Microgel synthesis}
We synthesised pNIPAM microgels using the free radical precipitation polymerisation method as previously described by Conley et al. \cite{Conley2016} N-isopropylacrylamide (Acros Organics, 99\%), NIPAM, is the monomeric unit which is recrystallised in hexane before use and N,N-methylene bis(acrylamide) (Sigma–Aldrich, 99\%), BIS, is the cross-linker. In addition, N-(3-aminopropyl) methacrylamide hydrochloride (Polysciences), APMA, is added as a co-monomer to incorporate free amine groups into the microgel network. The primary amines are used to fluorescently label the microgels by reacting with the succinimidyl ester groups of the dye AlexaFluor 647. 2,2-Azobis(2-methylpropionamidine) dihydrochloride (Sigma–Aldrich, 98\%), AAPH, is used to initiate the polymerisation.
First, in a three-neck round bottom flask, NIPAM (1.430 g) and BIS (0.029 g) are dissolved in 85 g \ce{H2O}. The reaction mixture is purged with nitrogen for 30 min before the temperature is raised to 70~\textdegree C. Next, 0.0365 g AAPH, previously dissolved in 5 g \ce{H2O}, is added to the reaction mixture. Five minutes after the initiator is added and the solution has started to turn white, 0.0198 g APMA dissolved in 10 g \ce{H2O} is introduced to the reaction using a syringe pump at an addition rate of 0.5 ml/min. The reaction mixture is kept at 70~\textdegree C for 4 h before cooling down rapidly in an ice bath. Following this protocol, inhomogeneous core-shell microgels are formed with APMA co-monomer only being present on the shell of the microgel network. A purification step follows to remove all unreacted monomers in the solution by centrifugation (3-4 times). We label the microgels using an excess amount of fluorescent dye Alexa 647 and leave them in an oscillation plate for 1 h before another purification step to remove all unreacted dye.
\\

\subsection{Super resolution microscopy}
To perform dSTORM experiments, we use a Nikon TiEclipse inverted microscope with an EMCCD camera  (Andor iXon Ultra 897)  and total internal reflection fluorescence (TIRF) arm to achieve highly inclined illumination and limit the fluorescence background noise. We use a continuous wave red laser, coherent Genesis MX-STM with 1000 mW output power at 639 nm, providing a single mode TEM00 Gaussian beam, horizontally polarised. The high power red laser enables fluorophores in their excited state through intersystem crossing to occupy the triplet state where they get trapped. A second laser (Toptica iBeam Smart) with 120 mW output at 405 nm, vertically polarized, is used to tune the blinking density. Both lasers are coupled into a single-mode  fiber (S405XP, Thorlabs) into the TIRF arm. The light is focused on the back aperture of a high numerical aperture and magnification objective (NA 1.49 and 100x magnification). With an extra zoom lens placed before the camera, the final pixel size is 110 nm. A dichroic filter with a wavelength of 700 nm and bandwidth of 50 nm is placed in the detection pathway (ET700/50, Chroma).\\
First,  we set the appropriate buffer conditions at 50 mM Cysteamine, and pH adjusted to 8. Second, we illuminate the sample at a wavelength of $\lambda=639$ nm using a high laser power (3.4 kW/cm$^{2}$) to bring the fluorophores to a metastable dark state. We employ a second laser at 405 nm to induce stochastic spare fluorescent light emission and control the blinking density by adjusting the laser power. We acquire 30000 - 60000 frames with an exposure time of 10 - 20 ms. Using \textit{Picasso} software, for each frame we localize individual blinking events that we then fit with a 2D Gaussian profile using the maximum likelihood estimation method (MLE).
Following this procedure, we generate a database that contains x-y positions, intensity, and localization precision (x,y). To reconstruct the entire super-resolution image we filter the list of localization and only retain localizations above a certain photon count threshold, with low ellipticity ($<$0.2) and with a sufficient localization precision in x and y ($<$ 0.2 pixel). We correct the reconstructed image's drift using a redundant cross-correlation algorithm and then render the image using the individual localization precision (iso) mode unless otherwise stated.

For the aging measurements, we took into account that the number of localization may vary with time, due to the fact that the buffer conditions change over long imaging time~\cite{Olivier2013}. In order to compare among the different waiting times, we then consider only the first 10000 blinking frames of the measurement, corroborating that the number of localizations per microgel (about 3000) were comparable for all cases.

\subsection{Light scattering measurements}
DLS measurements are performed using the commercial LS Spectrometer, 2D-DLS Pseudo cross correlation set-up (LS Instruments AG, Switzerland). We increase the temperature from 25 to 35~\textdegree C with $\Delta T$ = 1~\textdegree C as step size. Laser light with wavelength of 660 nm was used to perform the experiments at scattering angles of 40 - 80\textdegree~ with 5\textdegree~  step size. Three measurements of 40 seconds were obtained at every angle. The diffusion coefficients D and hydrodynamic radii $R_{H}^{DLS}$ were extracted from a multi angle analysis of the first order cumulant fit.

\subsection{Numerical Methods}
\subsubsection{Microgel Modeling}
We simulate the microgels assembly following the methods described in Refs.~\citenum{gnan2017silico,ninarello2019modeling}.
The assembly method is a two step process: first, {\it (i)} bi- and tetravalent patchy particles self-assemble inside a spherical cavity forming a fully-bonded disordered network; then, {\it (ii)}  the topology of the network gets fixed by replacing the patchy interactions with permanent bonds. 
For the fixed topology,  particles composing the microgel, which we will also refer to as monomers, interact with the Kremer-Grest bead-spring model~\cite{kremer1990dynamics}, i.e.  monomers overlap is avoided with the repulsive Weeks-Chandler-Andersen (WCA) potential~\cite{weeks1971wca}
\begin{equation}\label{eq:wca}
    V_{\text{WCA}}(r)=\begin{cases}
    4\epsilon \left[ \left( \frac{\sigma}{r} \right)^{12} - \left( \frac{\sigma}{r} \right)^{6}  \right] + \epsilon     & \text{ if } r\leq 2^{1/6}\sigma \\
    0 & \text{ otherwise } 
    \end{cases}
\end{equation}
while permanent bonding between monomers is ensured by adding the  Finite-Extensible-Nonlinear-Elastic (FENE) potential for bonded pairs~\cite{Warner1972FENE}
\begin{equation}\label{eq:fene}
    V_{\text{FENE}}(r)= -\epsilon k_{\text{F}} R_{0}^{2} \ln{\left[ 1 - \left( \frac{r}{R_{0}\sigma} \right)^{2}  \right]}   \text{ if } r\leq R_{0}\sigma
    \end{equation}
where $\sigma$ is the monomers diameter, $\epsilon$ the energy scale, $k_{\text{F}}=15$ the dimensionless spring constant, and $R_{0}=1.5$ the maximum bond extension.
The volume phase transition of the microgels is reproduced by adding an attractive solvophobic interaction $ V_{\alpha_{\text {mm}}}(r)$ among monomers~\cite{soddemann2001generic,verso2015simulation}. The attraction strength is controlled by the parameter $\alpha_{\text {mm}}$ which encodes the monomer-monomer effective attraction, modeling implicitly the reduction of monomer-solvent affinity when increasing the temperature:
\begin{equation} \label{eq:valpha}
    V_{\alpha_{\text {mm}}}(r)=\begin{cases}
    -\epsilon\alpha_{\text {mm}} & \text{ if } r\leq 2^{1/6}\sigma \\
    -\frac{1}{2}\epsilon\alpha_{\text {mm}} \left[ \cos{\left( \gamma (r/\sigma)^{2} + \beta \right) }  -1 \right] & \text{ if } 2^{1/6}\sigma < r \leq R_{0}\sigma\\
    0 & \text{ otherwise }
    \end{cases}
\end{equation}
with $\gamma = \pi (2.25 - 2^{1/3})^{-1}$ and $\beta = 2 \pi - 2.25 \gamma$. When $\alpha_{\text {mm}}=0$ no implicit-solvent attraction is added between monomers, reproducing good solvent conditions and a swollen microgel; instead, monomers attraction rises by increasing $\alpha_{\text {mm}}$, thus shrinking the microgel size and mimicking the worsening of the implicit solvent.

\subsubsection{Solid Surface Modeling and Monomers Anchoring}
To reproduce the behavior of the microgel close to a surface, we model the latter as two layers of wall particles, that are initially located on a compact square lattice of site $\sigma$ at the bottom of the simulation box; the separation between layers is $0.7\sigma$.
To avoid crystallization of the monomers close to the surface, the wall particles are randomly displaced from the lattice sites, including the direction perpendicular to the plane, following a Gaussian distribution with standard deviation $\sigma_{\text{sd}}=0.2$. The obtained layers are then subsequently fixed throughout the whole simulation runs.

Microgel monomers interact with wall-particles via the WCA potential eq.~\ref{eq:wca} and the $V_{\alpha_{\text {ms}}}$ potential, which is identical to  Eq.~\ref{eq:valpha}, but this time replacing the monomer-monomer attraction $\alpha_{\text {mm}}$  with the  
monomer-surface one,  $\alpha_{\text{ms}}$, now encoding the surface hydrophobicity.

\begin{figure}[ht]
\centering
  \includegraphics[width=3.33in]{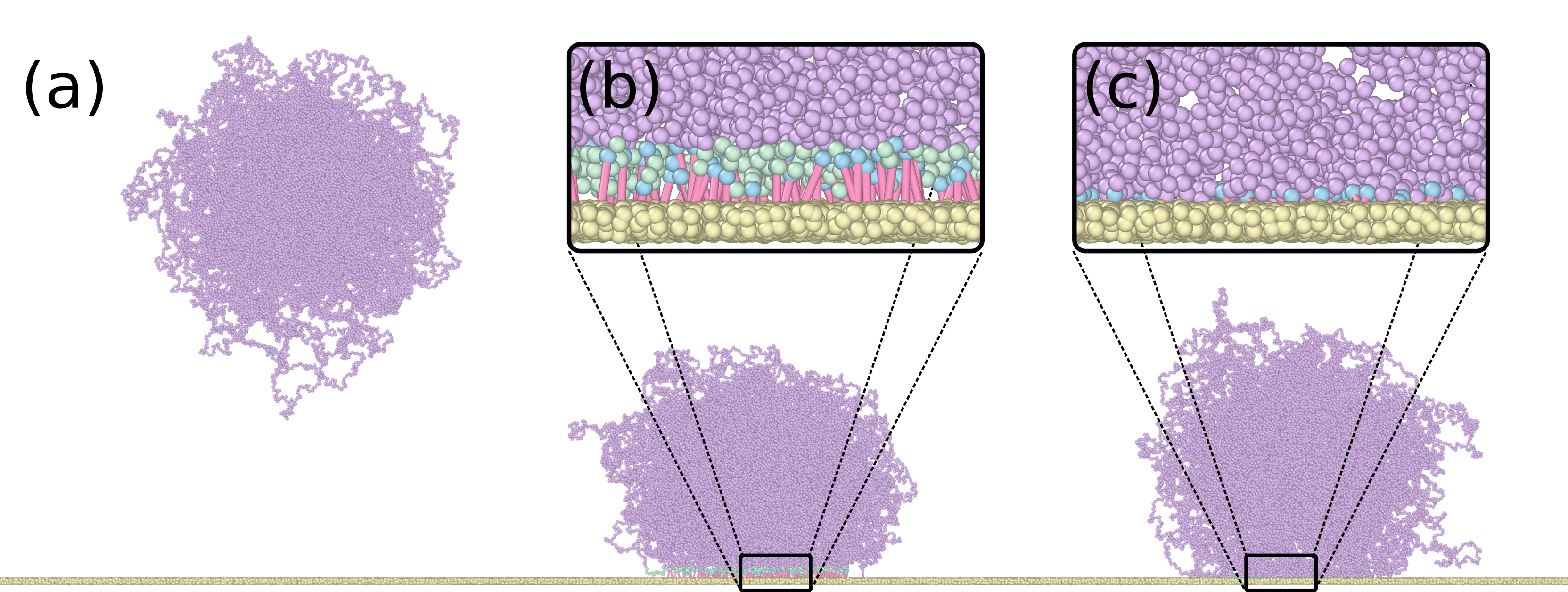}
  \caption{Bonding of the microgel to the hydrophilic surface: (a) the microgel in the bulk is pushed towards the wall, (b) when in contact, a list is made with all monomers at a distance to the wall smaller than $dz$ (monomers in green), from which some are randomly chosen (blue) and bonded to their closest wall-particle (pink bond), (c) the microgel is let to relax.}
 \label{fgr:mgel-to-wall}
\end{figure}

To mimic experimental conditions where the microgel is physically anchored to the wall, we also consider the case where permanent bonds between a few monomers and the surface particles are formed. This is obtained by the following procedure (illustrated in Fig.~\ref{fgr:mgel-to-wall} for the hydrophilic scenario $\alpha_{\text{ms}}=0$): {\it (i)} a swollen microgel (equilibrated in bulk at $\alpha_{\text{mm}}=0$) is pushed towards the wall; 
{\it (ii)} when it comes in contact with the surface, monomers with distance less than $dz=1.5\sigma$ from the upper layer of the wall are considered, and among them, {\it (iii)} $b$ monomers are randomly chosen and anchored to their closest wall-particle via a harmonic potential $V(r)=K(r-r_{0})^{2}$ with $K=15$ and $r_{0}=2^{1/6}\sigma$; finally, {\it (iv)} the microgel is let to relax to its equilibrium state. 
The procedure is then repeated for different surface $\alpha_{\text{ms}}$ conditions. We did it both for the hydrophilic $\alpha_{\text{ms}}=0$ and hydrophobic $\alpha_{\text{ms}}=0.9$ conditions, yielding different bonding patterns. Density profiles calculated with microgels anchored to a hydrophilic surface are comparable to experiments in all cases, except for the measurements at $T = $ 35\textdegree C close to a hydrophobic surface, where we found that the extension of the tail is better captured by simulations of microgels initially anchored to a hydrophobic surface (see Fig.~S15).
Another important parameter to take into account is the number of bonds $b$ that we should consider in the simulations. As mentioned in the Results Section and shown in Fig.~S11, for the hydrophilic surface we tried different values of $b$ and found that $b=25$ is the most similar to experimental data. For the hydrophobic case, we expect in experiments a much larger number of bonds due to the additional attraction to the surface in the anchoring procedure. For this reason, we performed all simulations (that take much longer, also due to the long aging regime) with a fixed value $b=200$, roughly one order of magnitude difference with respect to the hydrophilic case. This value was then found to be in rather good agreement with experiments in terms of the tail of the density profiles.

\subsubsection{Fluorophore Emulation}
Super-resolution experiments only detect the presence of fluorophores labeling the microgels. We thus need a way to go from the full monomer representation of the microgel to the case of only counting fluorophores in the calculation of the density profiles to mimic the partial labeling performed in the experimental synthesis.
In our simulation this is equivalent to choose a particular set of monomers that are considered to be the fluorophores. Since microgels are synthesized in such a way that these are mostly located in the surface, the monomer-to-fluorophore conversion process must take this into account. To this aim, we perform the fluorophore selection with the following procedure, illustrated in Fig.~\ref{fgr:radialdensityprofile}(d):
we allocate monomers to the \textit{fluorophore list} according to their distance $r$ from the microgels center of mass (CM) in a single equilibrated configuration by testing if  $r > \langle R_{g} \rangle \cdot P(\sigma_{\text{sd}},\mu=1)$, where $P(\sigma_{\text{sd}},\mu=1)$ is a random number taken from a Gaussian distribution of mean $\mu=1$ and standard deviation $\sigma_{\text{sd}}$.  The addition of such a Gaussian noise softens the definition of the interface and takes into account averaging over different equilibrium configurations as well as the fact that there could be mislocalization or loss of resolution in the experiments. Indeed, we find that $\sigma_{\text{sd}}$  increases with  temperature, due to the fact that it is more difficult to resolve the core-shell interface and the position of the fluorophores once the microgel collapses, as shown in Fig.~S8.  We find that $\sigma_{\text{sd}}=0.0$ and $0.3$ for  25, 30 ~\textdegree C. However, at high temperatures, when the microgel approaches a full collapse, the fluorophore description loses its validity. In particular, at 38 (35)~\textdegree C for the hydrophilic (hydrophobic) case, respectively, as discussed in the main text, the density profiles do not show a peak in the outer shell, and we resort to consider all the monomers to calculate the numerical 2D density profiles. Instead, for 35~\textdegree C and hydrophilic surface a mixture of the two approaches is needed, because fluctuations from microgel to microgel are large and we have almost an equal population of fully resolved and hollow profiles. We thus average them following experimental proportions, using $\sigma_{\text{sd}}=0.4$ for the fluorophore distribution.

\subsubsection{Simulation Parameters}
Molecular dynamics (MD) simulations of three independent realizations of the microgels with different topologies were performed using LAMMPS\cite{plimpton1995fast}. Microgels were assembled using oxDNA package\cite{rovigatti2015oxDNA} starting with $N = 42 000$ patchy particles, of which 1.5\% are tetravalent (crosslinkers) to match experimental conditions. It is important to note that the adopted $N$ yields the size of a monomer to be around 9~nm, as described in the text. This value is slightly larger than the estimated Kuhn length for PNIPAM~\cite{lopez2019swelling}, but we previously showed~\cite{ninarello2019modeling} that, upon increasing the number of monomers, the bead size approaches the correct value without qualitatively affecting the results.
The wall is made of two layers of 90000 wall-particles each. Hence, each system has about 222000 particles in total.

We used a Langevin thermostat with reduced temperature $T^{*}=k_{\text{B}}T/ \epsilon=1$, particles mass $m=1$,  and integration time $\delta t=0.002\sqrt{m\sigma^{2}/\epsilon}$. Wall-particles are kept fixed by not including them in the integration scheme. 
The length of the simulations changes for the different scenarios; we monitored the energy as to see when the system had thermalized, and then ran additional steps from where the density profiles were calculated. 
Simulations in the bulk and on a hydrophilic surface take around $1\times10^{6}$ steps to equilibrate. After this, we ran additional $15\times10^{6}$ steps. Density profiles are calculated from configurations in the last $5\times10^{6}$ steps of the simulation. Instead,  the evolution of microgels placed close to a hydrophobic surface is very slow. Hence, in this case we performed $100\times10^{6}$ steps. We calculated the profiles at different waiting times and found that their shape did not show significant changes (except at very large distances), and is similar to what reported in Fig.~\ref{fig:phobic-T25}.

The radial density profiles of individual equilibrated microgels were calculated as 
\begin{equation}\label{eq:rho_r}
\rho(r)=\left\langle  \frac{1}{N} \sum_{i=1}^{N} \delta \left( \left| \vec{r}_{i} - \vec{r}_{\text{CM}} \right| - r \right) \right\rangle
\end{equation}
where $\langle \cdot \rangle$ is the average over several configurations. When calculating the 2D profiles, $\vec{r}$ and $\vec{r}_{\text{CM}}$ do not include the component perpendicular to the surface. Observables were averaged over three independent microgel configurations in order to avoid singular topology characteristics that may occur.  Images of numerical microgels were created with ovito~\cite{Stukowski2010ovito}.

\begin{acknowledgement}
We thank G. Del Monte for help with the calculation of the hydrodynamic radius. X.S., R. R.-B., F.S. and E.Z. acknowledge funding from the European Union's Horizon 2020 research and innovation program under the Marie Sk\l odowska-Curie (ITN SUPERCOL, Grant Agreement 860914). This work has benefited from financial support from the Swiss National Science Foundation through the National Centre of Competence in Research 'Bio-Inspired Materials' and project number \# 149867 (F.S. and M.B.). 
\end{acknowledgement}

\begin{suppinfo}

Additional details on the sample preparation for dSTORM measurements, the image analysis, and the density profiles fit using the fuzzy sphere model for microgels on the hydrophilic surface; also additional simulation details regarding the 2D projected density profiles, the estimation of the fluorophore distribution, the comparison between bulk and hydrophilic surface results, the monomer-surface parameter and the anchoring of the microgel to the surfaces.

\end{suppinfo}

\bibliography{rsc.bib}

\begin{mcitethebibliography}{70}
\bibitem[foo()]{footnote}
Given that the error bar on the experimental $R_g$ is of the order of 10 nm, we
  choose the numerical value within the experimental interval, finding that the
  best agreement is at its lowest end. \relax
\mciteBstWouldAddEndPunctfalse
\mciteSetBstMidEndSepPunct{\mcitedefaultmidpunct}
{}{\mcitedefaultseppunct}\relax
\EndOfBibitem
\end{mcitethebibliography}

\end{document}


\title{Supplementary Information for: \\ Probing temperature-responsivity of microgels and its interplay with a solid surface by superresolution microscopy and numerical simulations}

\author{Xhorxhina Shaulli}
\altaffiliation{Equal first author}
\affiliation{Department of Physics, University of Fribourg, Chemin du Musée 3, 1700, Fribourg, Switzerland}
\author{Rodrigo Rivas-Barbosa}
\altaffiliation{Equal first author}
\affiliation{Department of Physics, Sapienza University of Rome, Piazzale Aldo Moro 2, 00185 Roma, Italy}
\author{Maxime Jolisse Bergman}
\affiliation{Department of Physics, University of Fribourg, Chemin du Musée 3, 1700, Fribourg, Switzerland}
\author{Chi Zhang}
\affiliation{Department of Physics, University of Fribourg, Chemin du Musée 3, 1700, Fribourg, Switzerland}
\author{Nicoletta Gnan}
\affiliation{CNR Institute of Complex Systems, Uos Sapienza, Piazzale Aldo Moro 2, 00185, Roma, Italy}
\affiliation{Department of Physics, Sapienza University of Rome, Piazzale Aldo Moro 2, 00185 Roma, Italy}
\author{Frank Scheffold}
\altaffiliation{Equal senior author}
\email{frank.scheffold@unifr.ch}
\affiliation{Department of Physics, University of Fribourg, Chemin du Musée 3, 1700, Fribourg, Switzerland}
\author{Emanuela Zaccarelli}
\altaffiliation{Equal senior author}
\email{emanuela.zaccarelli@cnr.it}
\affiliation{CNR Institute of Complex Systems, Uos Sapienza, Piazzale Aldo Moro 2, 00185, Roma, Italy}
\affiliation{Department of Physics, Sapienza University of Rome, Piazzale Aldo Moro 2, 00185 Roma, Italy}

\renewcommand{\thefigure}{S\arabic{figure}}

\maketitle

\section{Sample preparation for dSTORM }
\begin{figure}[ht]
\centering
  \includegraphics[height=4cm]{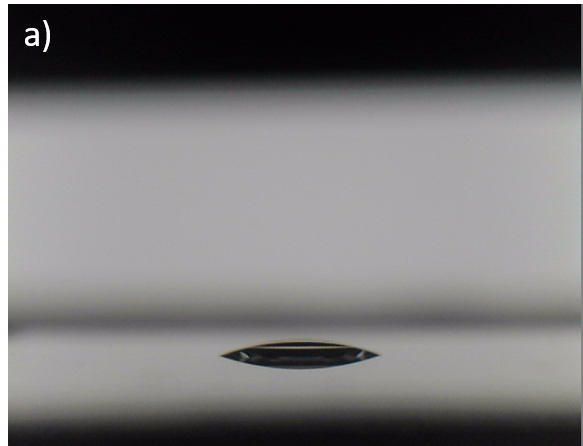}
  \includegraphics[height=4cm]{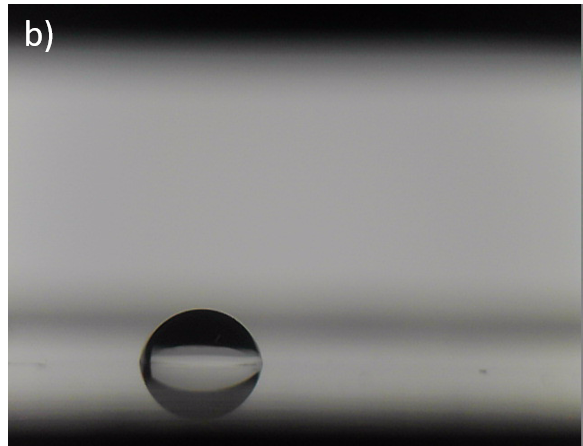}
  \caption{ \label{fig:contactangle}
  Images of 4~\textmu L water droplets on the different surfaces for the determination of the contact angle. (a) Contact angle measured on hydrophilic surface, angle: $ < 20 $\degree. (b) Contact angle measured on hydrophobic surface, angle: $ >80 $\degree. }
\end{figure}

 To prepare the hydrophilic surface, the coverslip is previously treated with KOH 3~M and sonicated for 10 min followed by exposure for additional 10 min on a UV ozone oven. After treatment the contact angle was measured, as shown in Fig.~\ref{fig:contactangle}(a). For the hydrophobic surface, the coverslip was first cleaned with KOH 3~M and then exposed overnight to 0.1 mL Hexamethyldisilazane (HMDS). The contact angle was measured after treatment, as reported in Fig.~\ref{fig:contactangle}(b). In order to perform a dSTORM experiment on individual microgels, the particles have to be spaced from one another and fixed on the surface. To this aim, we dilute the microgel solution and place around 7~\textmu L between two coverslips in order to create a thin layer and put it to dry at 55~\degree C shortly. The immobilized particles are then resuspended in a buffer solution containing $\beta$-Mercaptoethylamine (Sigma-Aldrich) at concentration 50~mM and pH is adjusted to 8 using HCL 0.1~M. Commonly the buffer solution may require also an oxygen scavenger system, but to keep the system as simple as possible we avoid this and simply make sure to seal the coverslip completely with  Picodent glue (Twinsil).

\section{Image Analysis }
The raw data are extracted from the microscope as files with .hdf5 extension, which can be read with several different open source software to reconstruct the super resolved image. In the present case we use Picasso software for processing and post processing of the data as it is fast and easy to use.\cite{Schnitzbauer2017} Picasso has several components. The first one we use to reconstruct the image in Localize. The following  camera settings are added  in the main window: 
photo-electrons per A/D count $=$ 5 (Sensitivity), base level $=$ 55, EM gain $=$300, pixel size $=$ 110~nm. In each frame the single molecule spots are identified and fitted using the Maximum Likelihood Estimation (MLE). The spots are found by considering the net gradient towards the bright center. Then a box of 7 pixels is drawn around it and the exact center is fitted. This is a localisation. For each spot fitted with MLE, the Cramer-Rao lower bound (CRLB) is estimated, the square root of  which gives us the localization precision.
The second step is image post processing. Here, we first use the Filter component. We filter the data excluding all the selected spots that have low amount of photon counts or very high photon counts which probably come from more than one fluorophore blinking at the same time, elipticity higher than 0.2 and localization precision higher than 0.20~pixel in x,y.
The last step is rendering. Each localization is shown in Render, with possible smoothing effects. We use individual localization precision in all our images presented in the paper unless otherwise stated. A very important post processing step is the drift correction. In Render we use the localization-events-based drift correction where images according to their appearance in the movie are split into segments of 1000. Further post processing tools like Average and Pick are used when stated.
For each experimental data set only particles with enough localisations are kept. For that we pick in render all the particles that we are satisfied with and dismiss the others. All the picked particles are assigned random colors when reopened in Render. Reconstructed dSTORM images for microgels sitting on hydrophilic and hydrophobic surface are shown in In Fig.~\ref{fgr:xs3uv} and Fig.~\ref{fgr:xs3H} respectively.\\
The histograms of the $R_g$ for all experimental data sets are shown in Fig~\ref{fgr:rg_all}.
\begin{figure}[ht]
\centering
  \includegraphics[width=\textwidth]{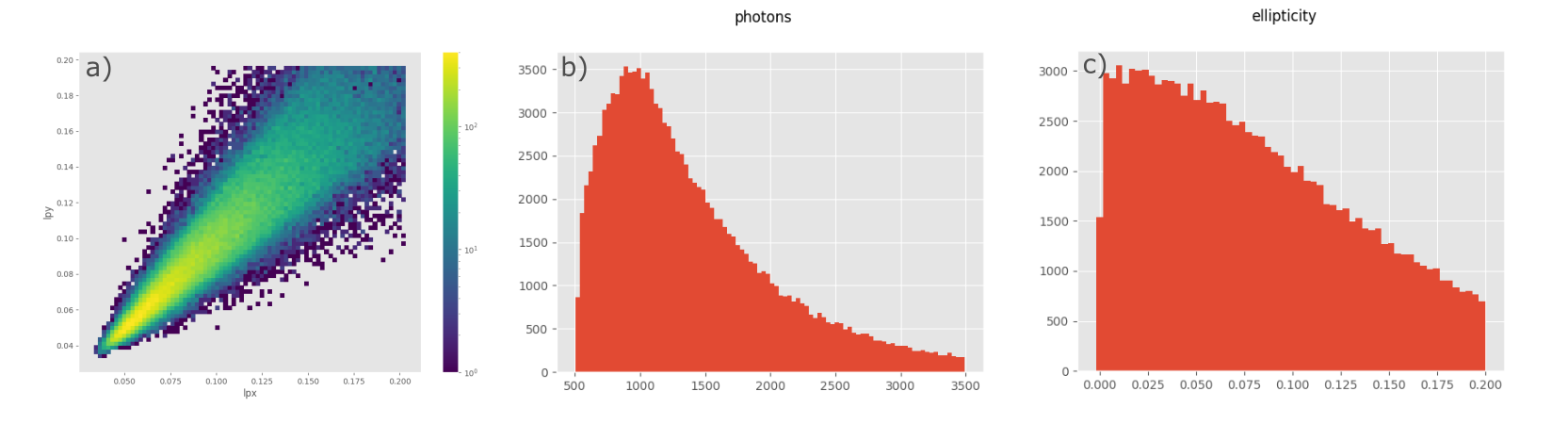}
  \caption{a) 2D histogram of the localization precision in x and y direction, in camera pixels, as estimated by the Cramer-Rao Lower Bound of the Maximum Likelihood fit. b) Filtered photon count distribution from one experimental data set. c) Filtered ellipticity distribution from one experimental data set.  }
 \label{fgr:example2}
\end{figure}

\begin{figure}[ht]
\centering
  \includegraphics[width=\textwidth]{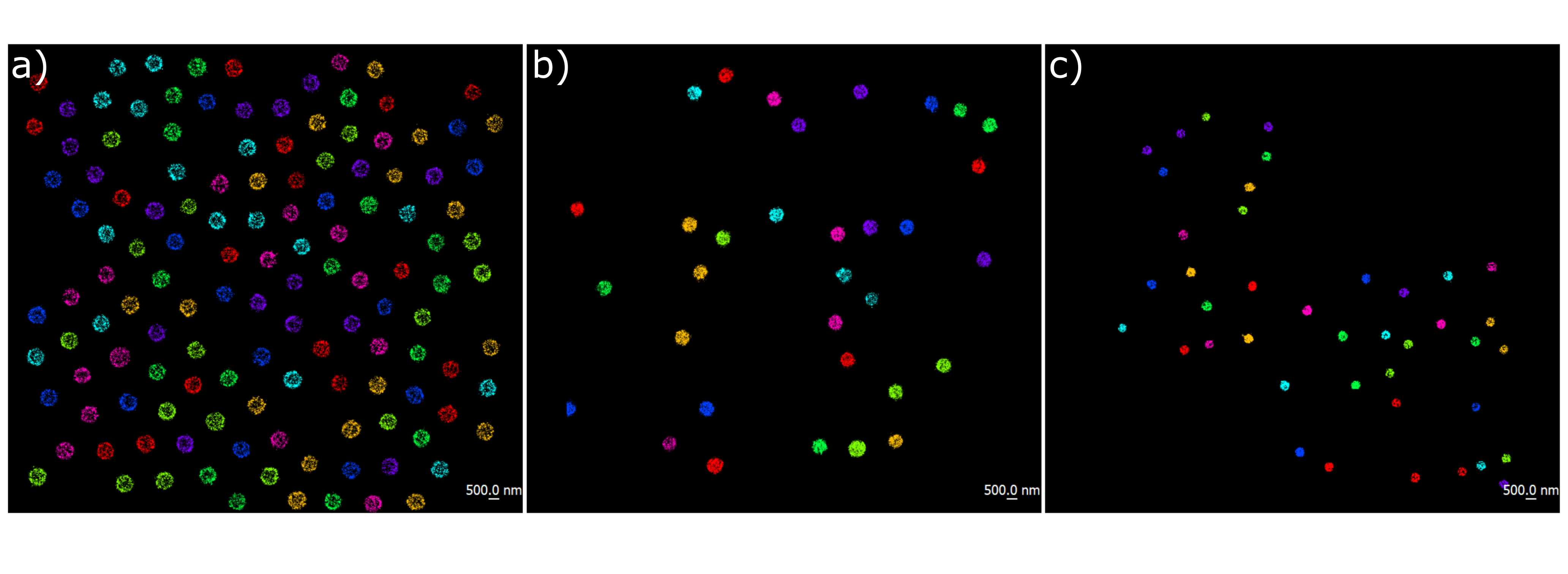}
  \caption{dSTORM images of microgels on hydrophilic interface for different temperatures a) 25\degree C, b) 30\degree C, c) 35\degree C. }
 \label{fgr:xs3uv}
\end{figure}
 
\begin{figure}[ht]
\centering
  \includegraphics[width=\textwidth]{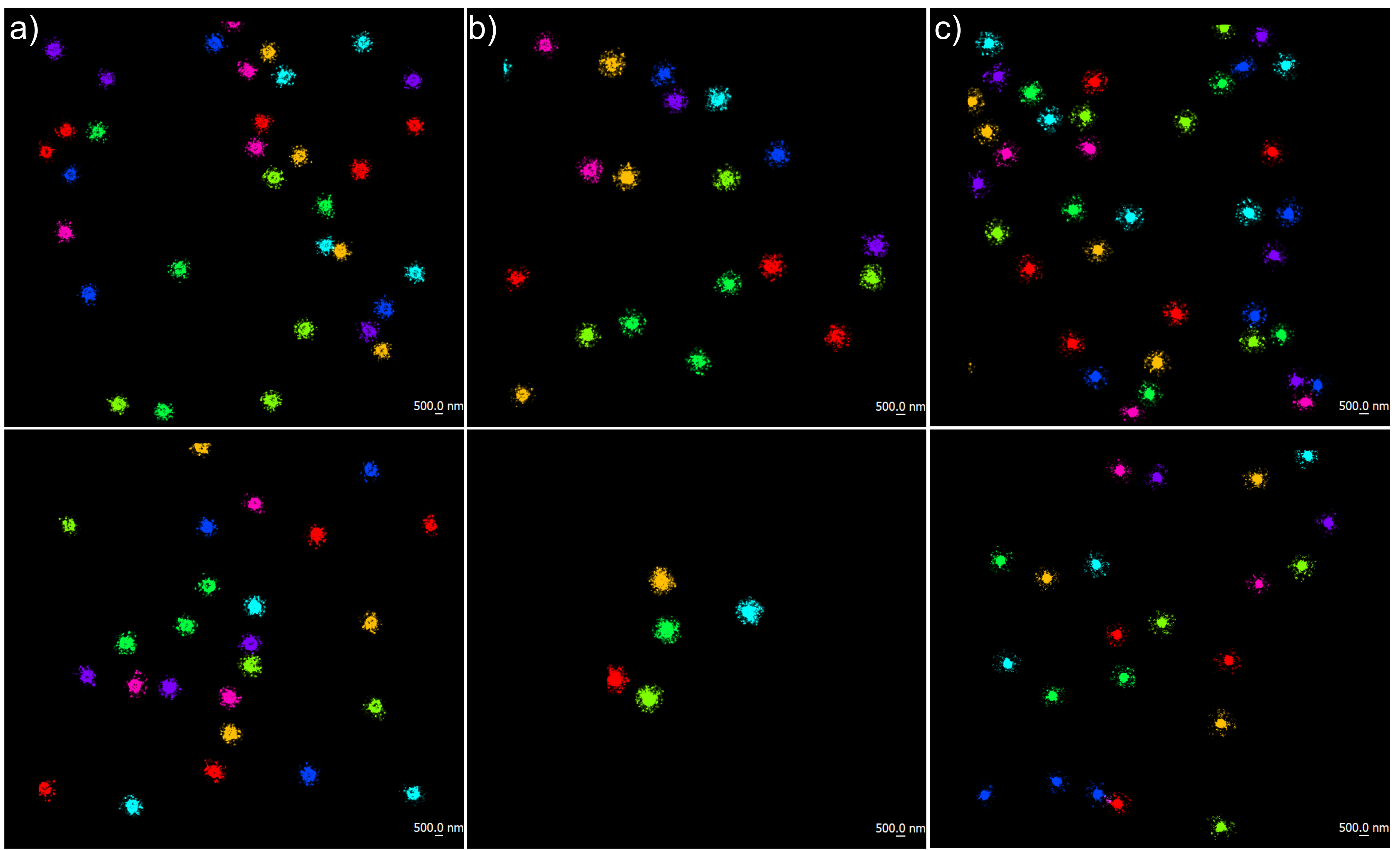}
  \caption{dSTORM images of microgels on hydrophobic interface for different temperatures a) 25\degree C, b) 30\degree C, c) 35\degree C. 
  }
 \label{fgr:xs3H}
\end{figure}

\begin{figure}[ht]
\centering
  \includegraphics[width=\textwidth]{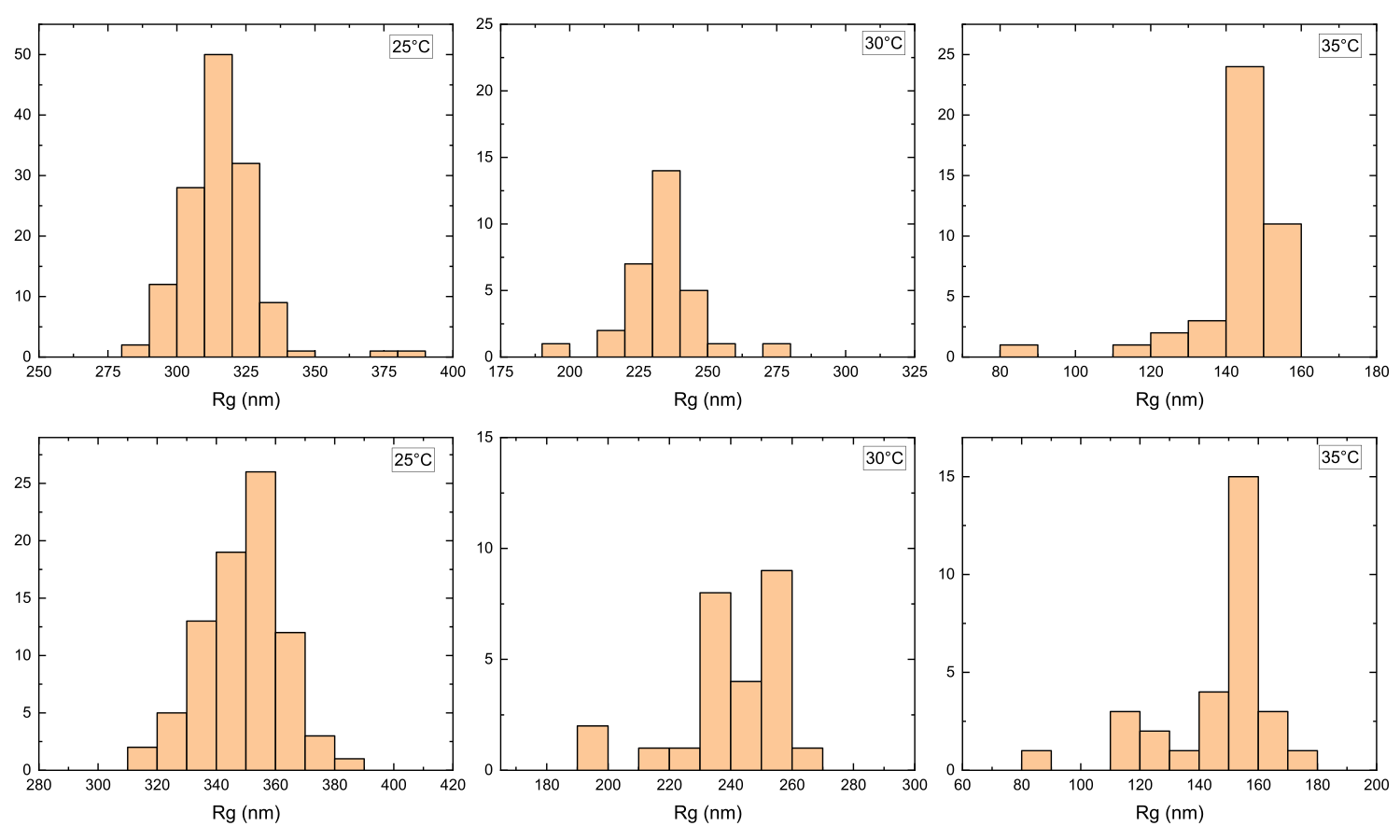}
  \caption{Experimental 2D $R_g$ Histograms for microgels at 25, 30 and 35 \textdegree C. Hydrophilic interface (top panel) and hydrophobic interface (bottom panel)}
 \label{fgr:rg_all}
\end{figure}

\clearpage
\section{Fit model - Hydrophilic interface }

\begin{figure}[ht]
\centering
  \includegraphics[width=0.6\textwidth]{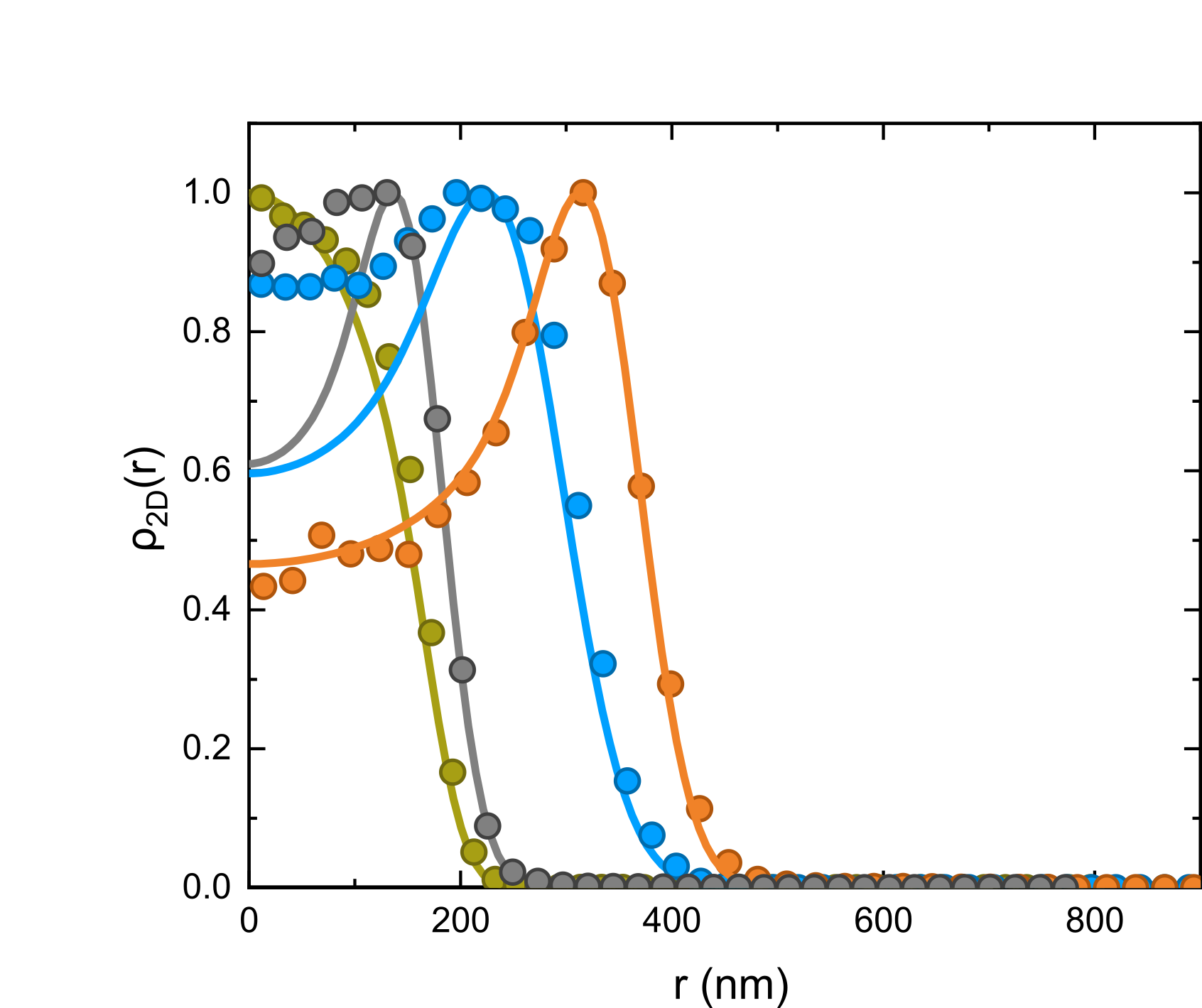}
  \caption{dSTORM analysis of the microgel radial density profiles illustrating a measured 2D profile (symbols). Fit with a Gaussian ring or a fuzzy sphere with a core radius R, and a
shell parameter $\sigma$ (lines). Microgels at 25\textdegree C (red), 30\textdegree C (blue),  35\textdegree C (gray) and 38\textdegree C (green). The plots are normalized by setting the peak value to one. }
 \label{fgr:fit_maxim}
\end{figure}

The 2D projected density profiles as obtained from dSTORM experiments have been compared to calculated density profiles. The microgels adsorbed to a hydrophilic surface retain their original, spherical swollen shape. In order to emulate the radially decaying density of the microgel under swollen conditions, the classic fuzzy sphere model predicts~\cite{Stieger2004,Conley2016}. 
\begin{equation} \rho_\textrm{3D}(X,Y,Z)= \rho_\textrm{3D}(r)=\text{erfc}\left[\frac{r-R}{\sqrt{2}\sigma_{surf}}\right]/2 \label{erfprof}\end{equation}, here normalized to $\rho_\textrm{3D}(0)=1$ and with $r=\sqrt{X^2+Y^2+Z^2}$ (alternatively the profile can be normalized for a constant microgel mass, see reference~\cite{Conley2016}). Because the microgels contain fluorophores predominately in their outer shell, their denser core can be considered `invisible' in the microscopy videos. We could adapt the fuzzy sphere model to account for the fluorophore profile contained within the microgel using the following expression $2 \rho_\textrm{3D}(r)=\text{erfc}\left[(r-R)/(\sqrt{2}\sigma)\right]\tanh\left[(r-R_\text{hole})/(\sqrt{2}\sigma_\text{hole})\right]$ introducing additional fit-parameters $R_\text{hole}<R$ and $\sigma_\text{hole}$. In practice we find that a Gaussian ring profile yields a fit of similar quality   
\begin{equation}
\rho_\textrm{3D}(r)=\frac{1}{\sigma_{surf}\sqrt{2\pi}} \exp\left[-\left(\frac{r-R}{\sqrt{2}\sigma_{surf}}\right)^2\right].
\end{equation}
where $R$ designates core radius and $\sigma_{surf}$  indicates how fuzzy the profile is. The model 3D object is thus hollow but the transition between shell and core is not sharp. In addition, the radial decay of the microgel is preserved. In order to confidently compare to the experimental density profiles, the 3D density profile is projected onto 2D by integrating over $Z$. Next, the projected 2D density profile is convolved with Gaussian 2D filter using the experimental resolution in order to simulate the smearing of the experiment. The theoretical density profiles have been fitted manually to the experimental data. As the hollow core is not the main focus of our study, we concentrate less on fitting the first part of the curve which represents the smooth transition between core and shell, and more on precisely fitting the shell peak and the tail which is sufficient to give us the correct $R_{tot}$ as shown in fig.~\ref{fgr:fit_maxim}. We finally note that at 38\textdegree C, the presence of the hollow core is not visible any more and we fit the data using Eq.~\eqref{erfprof} (with $\sigma_{surf}\sim 0$).

\clearpage
\section{Additional Details on Simulations}

\subsection{Comparison of 2D and 3D density profiles}

\begin{figure}[ht]
\centering
  \includegraphics[width=\textwidth]{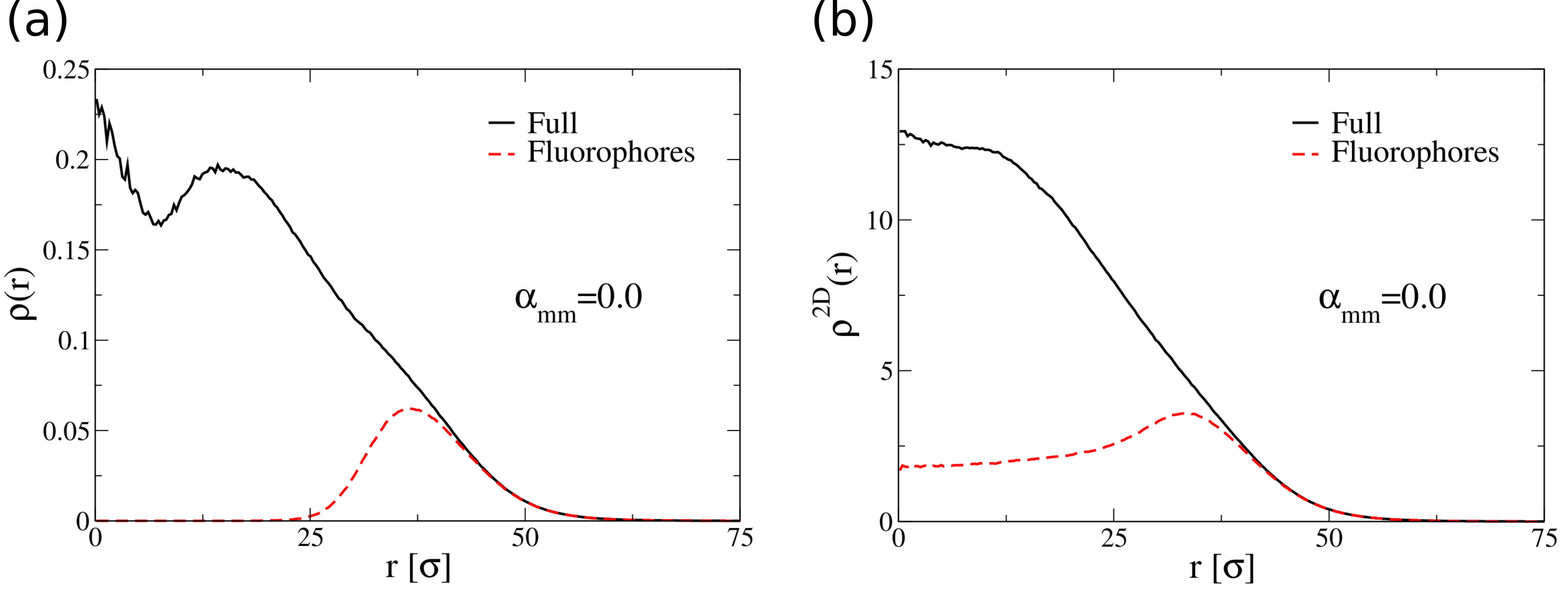}
  \caption{(a) 3D and (b) 2D density profiles of a fully (solid lines) and partially labeled (dashed lines) swollen microgel in the bulk.}
 \label{fig:dp2d-3d}
\end{figure}

In the main text we compare experimental and numerical 2D density profiles. For completeness, here we also report the corresponding 3D density profiles, calculated from the simulations, and we also differentiate between fully and partially labeled microgels. Fig.~\ref{fig:dp2d-3d} shows the (a) 3D and (b) 2D density profiles of fully and partially labeled swollen microgels ($\alpha_{\text{mm}}=0.0$) in the bulk. Unlike the 3D case, the partially labeled 2D profile is non-zero at short distances, due to the projection of fluorophores into the plane. The oscillations that occur at short distances in the 3D full profile are due to the fact that here we are reporting data for one microgel realization. To this aim, it is worth noting that in the manuscript results are always averaged over three independent topologies. However, the noise at small distances even for one realization is partially removed from the 2D projection, as visible in Fig.~\ref{fig:dp2d-3d}(b).

\FloatBarrier
\subsection{Estimating the fluorophore location within the microgels}
\begin{figure}[ht]
\centering
  \includegraphics[width=\textwidth]{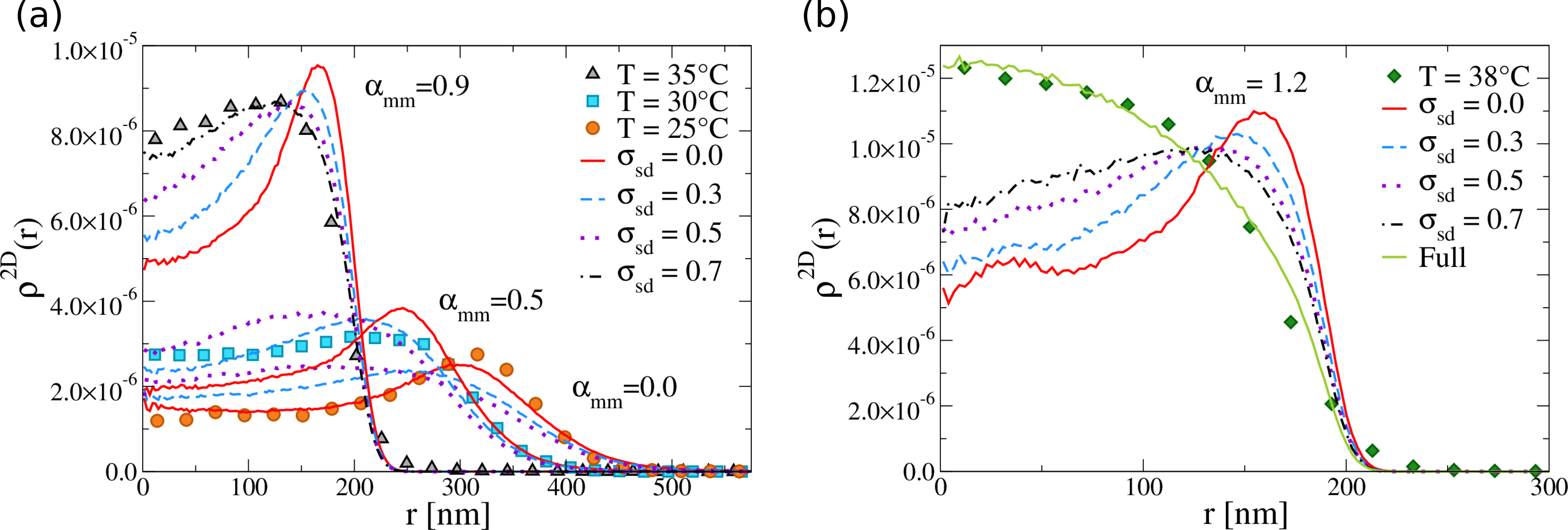}
  \caption{Experimental 2D density profiles (symbols) compared with numerical data with varying noise:  $\sigma_{\text{sd}}=0.0$ (red solid curves), 0.3 (blue dashed curves),  0.5 (purple dotted curves), and 0.7 (black dash-dotted curves) values. Panel (b) includes the numerical full microgel density profile (solid green) for the 38~\textdegree C temperature.  }
 \label{fig:dp_noise-sd}
\end{figure}

The 2D experimental and numerical profiles for the hydrophilic/bulk scenario at $T~=~25\degree$C agree very well without added noise.  
However, at higher temperature, the same does not hold, as shown in Fig.~\ref{fig:dp_noise-sd}, with deviations increasing with temperature.
To this aim, we adopt a Gaussian noise to add to the fluorophore location, as discussed in the main text, with standard deviation $\sigma_{\text{sd}}$. The results for different values of $\sigma_{\text{sd}}$ are also reported in Fig.~\ref{fig:dp_noise-sd}. Simulation profiles are in best agreement with experimental data close to the hydrophilic surface when the noise parameter is $\sigma_{\text{sd}}=0.0, 0.3, 0.7$ for $T = 25, 30, 35 \degree$C respectively. 
nstead, at $T = 38~\degree$C the experimental data cannot be fitted even with large noise, as shown in Fig.~\ref{fig:dp_noise-sd}(b), but as explained in the main text we find that they can be fitted by considering the full microgel profile (green solid line).  This is reinforced by the fact  that the microgel is now fully collapsed at this temperature and the 2D projection sees the microgel as a full, rather than hollow, object.

\begin{figure}[ht]
\centering
  \includegraphics[width=0.65\textwidth]{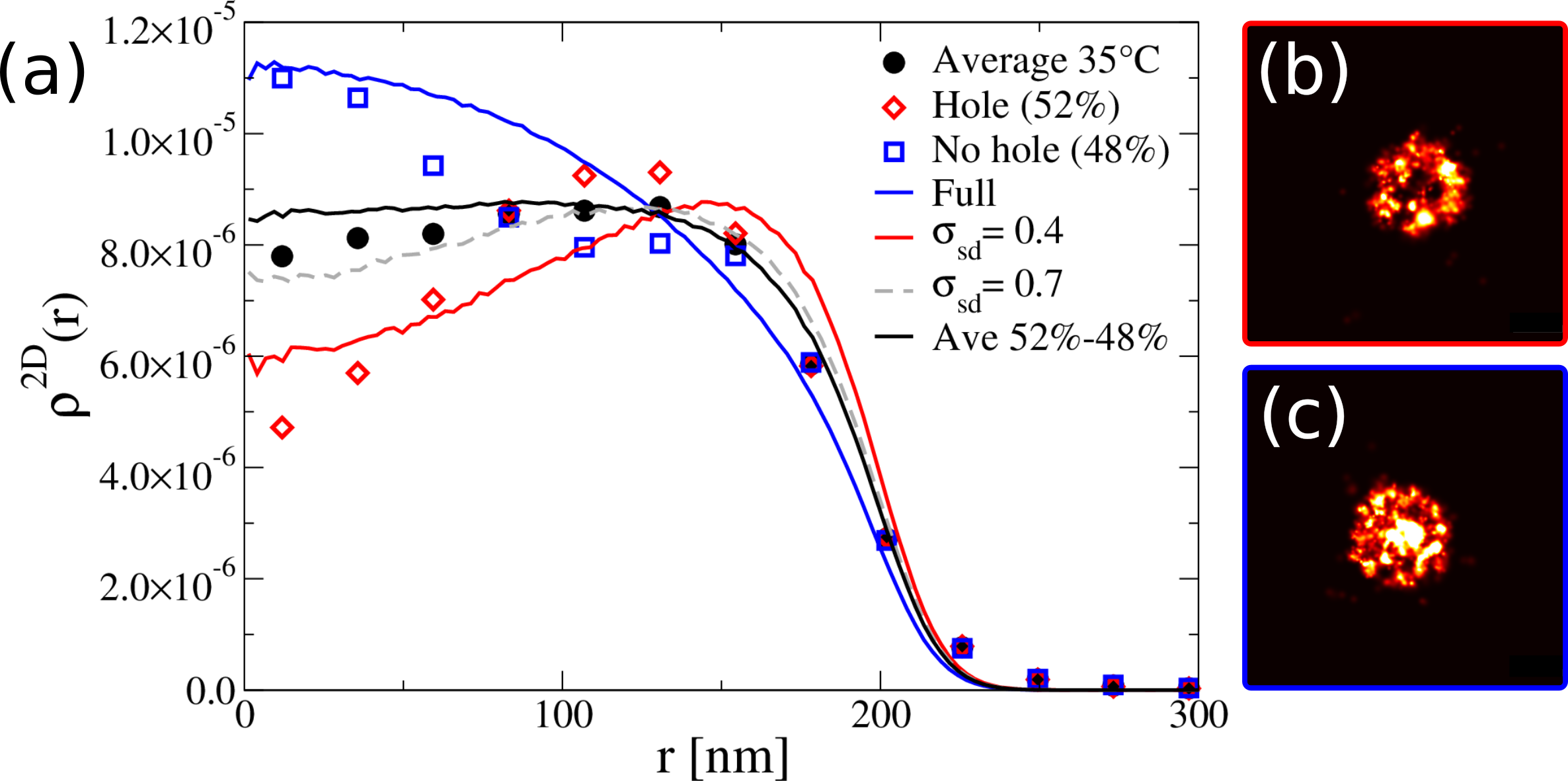}
  \caption{ Experimental 2D density profiles of averaged microgels with projected central hole (red diamonds, 52\% of all microgels), no hole (blue squares, 48\%), and total average (gray filled circles, 100\%); and numerical profiles (lines) fitting respectively. The experimental averaged profile can be fitted with both the averaged profile of the hole and no hole fittings (solid black line) and the noise $\sigma_{\text{sd}}=0.7$ profile (gray dashed line). Side panels: snapshots showing a microgel (b) with and (c) without a hole. }
 \label{fig:t35Ccomponents}
\end{figure}

Taking into account this feature, we also reanalyze the situation at T=35~\textdegree C, for which we find that a large numerical noise ($\sigma_{\text{sd}}=0.7$) is needed to fit the experimental profile.  We thus look deeper into each individual experimental profile at this temperature and find that 52\% of the microgels display a central 2D projected ``hole'' while the remaining 48\% do not show this.  Representative images of these situations are reported in Fig.~\ref{fig:t35Ccomponents} (b) and (c), respectively. Next, after classifying the microgels in ``hole'' or ``no hole'' groups, we fit the averaged experimental profiles of each group separately, as shown in Fig.~\ref{fig:t35Ccomponents}(a). While the ``no hole'' profile can be fitted with that of a full microgel,  the ``hole'' averaged profile can be described by the fluorophore ditribution with added noise $\sigma_{\text{sd}}=0.4$. We then calculate the weigthed average of ``hole'' and ``no hole'' profiles (black solid line) with the experimental profile, averaged over all analyzed microgels at this temperature (black circles). This is also compared with the previously found $\sigma_{\text{sd}}=0.7$ fit (dashed gray line) and we can conclude that both descriptions agree quite well with the experimental data. Given that the experimental data clearly show progressive filling of the hole upon increasing temperature, we thus adopt the description with the two populations of full and hollow microgels in the main text. Hence, the numerical description of the data in the main text Fig.~2 is the one corresponding to the black solid line in Fig.~\ref{fig:t35Ccomponents}(a). This allows us to use a full microgel profile to fit experiments where no hole is visible, which is the case for T=38~\textdegree C.

\begin{figure}[ht]
\centering
  \includegraphics[width=0.65\textwidth]{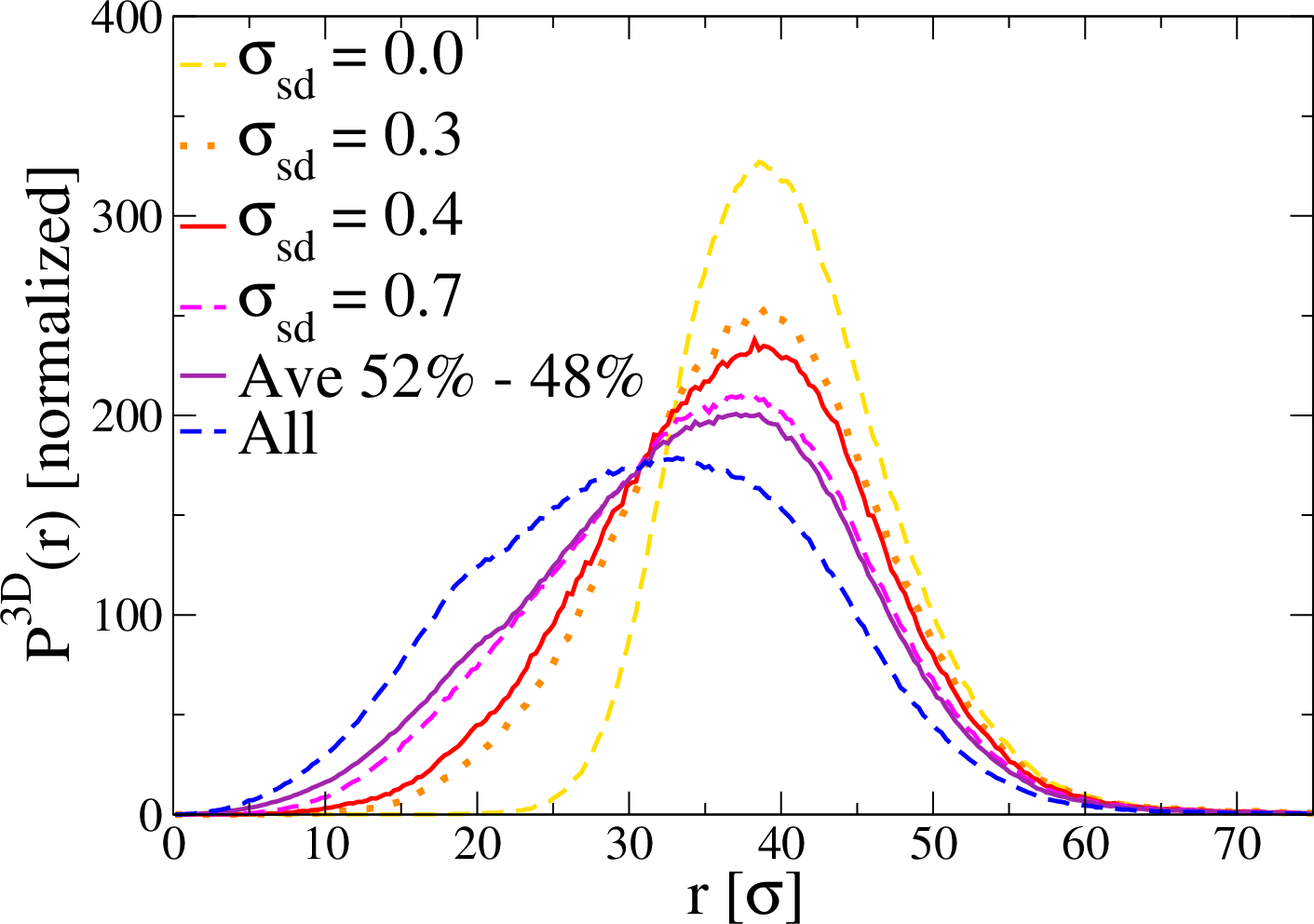}
  \caption{Averaged radial distributions of the fluorophore lists used to calculate the density profiles. The yellow dashed, orange dotted and pink dashed curves are the distributions calculated from the lists to fit the measurements at 25, 30 and 35~\textdegree C respectively, the blue curve is the distribution of the full microgel, used to fit the 38\textdegree C data.  Additionally, the purple  solid line is made from 52\% the red distribution plus 48\% the blue one, resulting similar to the pink dashed curve. The total microgel  average $R_{g}$, i.e., seen all monomers is $33.24 \cdot \sigma$.}
 \label{fig:hst3d_norm}
\end{figure}

In order to better explain this feature, we report in Fig.~\ref{fig:hst3d_norm} the averaged radial distribution of the different fluorophores used to calculate the density profiles when mapped to the swollen state $\alpha = 0.0$; in other words, for a microgel at $\alpha = 0.0$, we calculated the average radial distribution of the fluorophores used to fit the experimental data at the different temperatures. We also add the whole monomer distribution for comparison. Data are normalized to the same number.

It is clear that when increasing the noise $\sigma_{\text{sd}}$ the distribution broadens and shifts to shorter distances, meaning that when fitting higher temperatures, the fluorophores are seen to be closer the microgels core. However, the addition of the noise is not able to fully reach the limiting full profile case, as shown in  Fig.~\ref{fig:hst3d_norm}, where a small bump in the distribution occurs at small distance, probably due to the presence of the core, which is visible both in the full profile and the one where we mix full and fluorophore profiles (with noise $\sigma_{\text{sd}}=0.4$) as to describe the experimental data at 35~\textdegree C. Incidentally, this situation is very similar to the fluorophore profiles with noise $\sigma_{\text{sd}}=0.7$, also fitting such data, with the exception of the previously discussed bump.
Finally, it is important to note that, in the case of a hydrophobic surface, all monomers become visible at a lower temperature, namely 35~\textdegree C, probably because the full collapse is anticipated by the presence of the attractive surface, so that also in this case, we fit the experimental data with the full microgel profile and not with the fluorophore list (see also Fig.~\ref{fig:rebond} and related discussion below). These results altogether suggest that, for fully collapsed microgels, the fluorophores appear to be everywhere within the particle, due to the 2D projection of the data.  Of course, these considerations about the numerical fitting procedure only apply to the case of partially-labelled microgels and do not affect the main findings of the manuscript, but suggest that for a better and easier comparison to simulations, the dSTORM experiments should be better performed on fully labelled microgels.

\FloatBarrier
\subsection{Bulk vs hydrophilic comparison}
\begin{figure}[ht]
\centering
  \includegraphics[width=\textwidth]{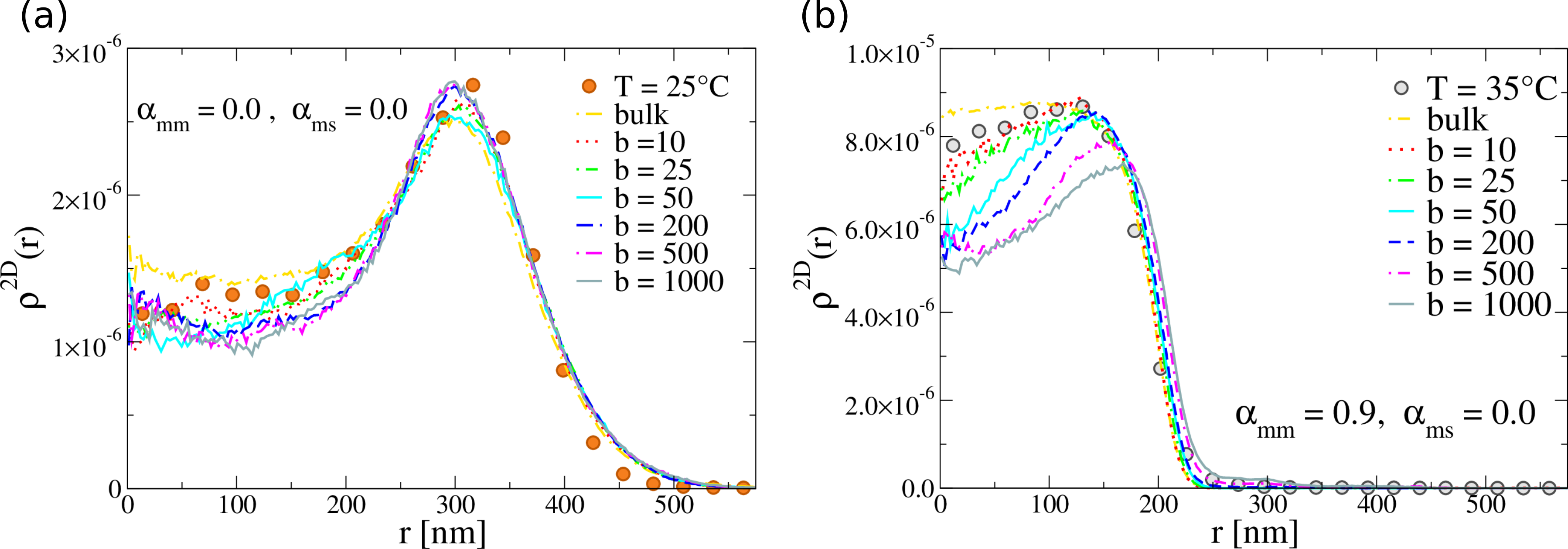}
  \caption{Experimental density profiles (symbols)  and numerical density profiles of a bonded
microgels anchored in a swollen state (lines) to a hydrophilic surface using different $b$ number of bonds (a) at 25~\textdegree C and (b)  35~\textdegree C. }
 \label{fig:dp_bnumber}
\end{figure}

We studied the effect of anchoring the microgel to a hydrophilic surface, following the anchoring procedure explained in the Methods. 
 To this aim, we repeat the anchoring procedure changing the number of surface-bonded monomers with the values $b=10, 25, 50, 200, 500, 1000$, in order to explore the influence of surface binding to the density profiles. After anchoring the microgel to the surface and letting the system to equilibrate, we perform simulations in hydrophilic conditions, i.e. $\alpha _{\text{ms}}=0.0$, for both swollen  $(\alpha_{\text{mm}}=0.0$) and collapsed ($\alpha_{\text{mm}}=0.9$) states.
Figs.~\ref{fig:dp_bnumber} (a) and (b) show the 2D density profiles at $\alpha_{\text{mm}}=0.0,0.9$ respectively. While in the swollen state (Fig.~\ref{fig:dp_bnumber}(a)) the profiles are indistinguishable for all values of $b$, in the collapsed state (Fig.~\ref{fig:dp_bnumber}(b)) some differences in the profiles can be seen, although the overall shape is similar for all cases. In particular, we find that the microgel extension gets  longer when increasing $b$ at the cost of a less dense core. This is due to an increase number of anchored monomers away from the 2D projection of the center of mass: these monomers are restrained from collapsing with the rest of the microgel.
Focusing on the short distance profiles in the collapsed state, we find that the optimum value of bonded monomers should be lower than $b < 50 $, meaning a number of anchored monomers below 0.1\%. Overall, we find that the number of anchored monomers for microgels on hydrophilic surfaces influences the density profiles only at temperatures above the VPT.

\begin{figure}[ht]
\centering
  \includegraphics[width=0.65\textwidth]{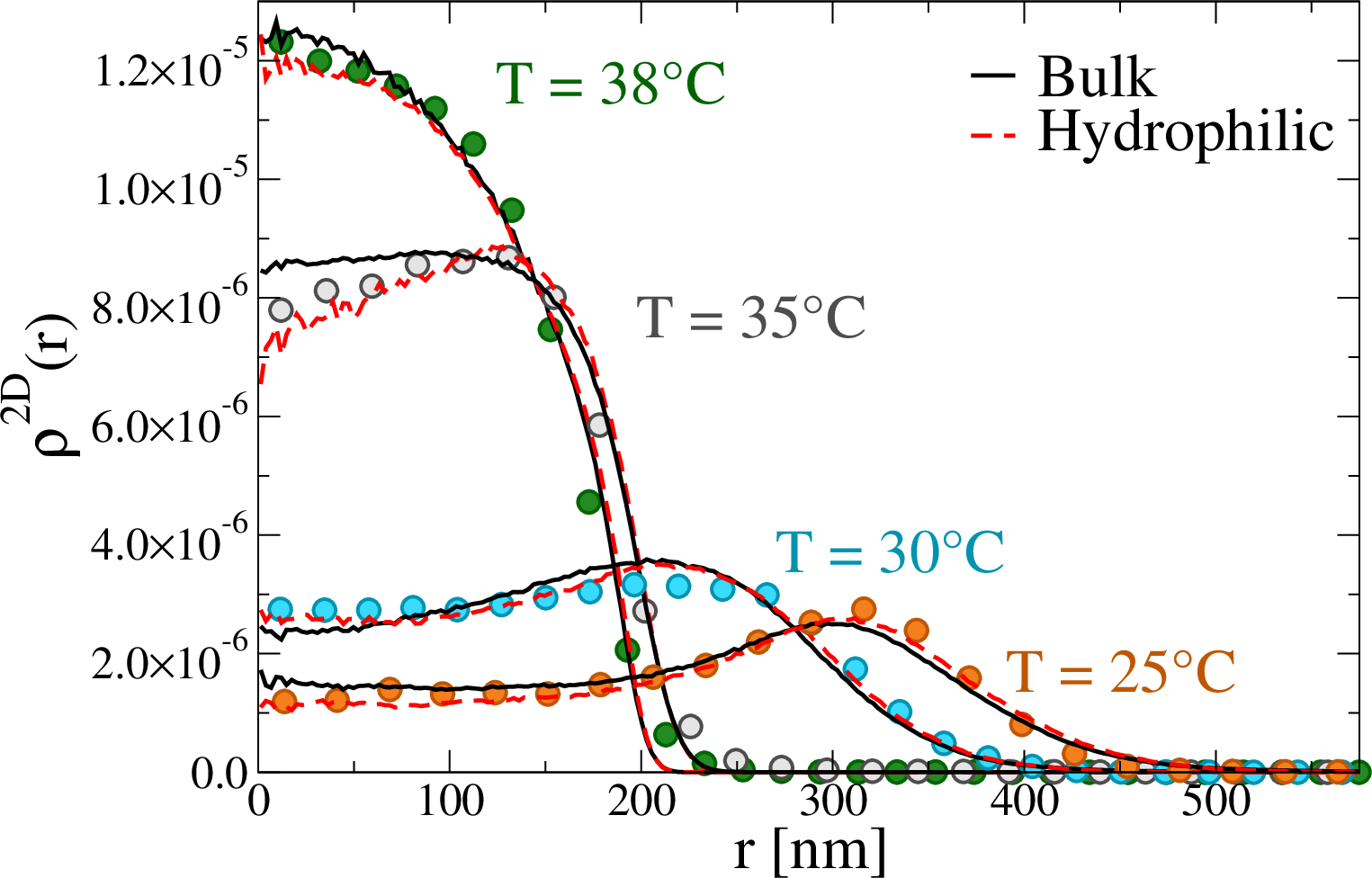}
  \caption{Experimental 2D density profiles (symbols) compared with numerical profiles of microgels in the bulk (solid black line) and anchored to a hydrophilic surface (red dashed line) with $b=25$ bonds at 25, 30, 35 and 38~\textdegree C.}
 \label{fig:bulk-philic}
\end{figure}

Based on the previous results, and to give a complete comparison between the bulk and hydrophilic scenarios, we simulated at all four temperatures ($\alpha$'s) microgels anchored with $b=25$ bonds.
Since we expect no attraction between the microgel and the surface, the interaction between monomers (other than the bonded monomers) and wall particles is exclusively repulsive via the WCA potential. Fig.~\ref{fig:bulk-philic} compares the calculated density profiles of hydrophilic wall-anchored microgels with those simulated in bulk and measured experimentally. The data are in very good agreement at all temperatures.

\FloatBarrier
\subsection{Selection of the monomer-surface attraction parameter }

\begin{figure}[ht]
\centering
\includegraphics[width=\textwidth]{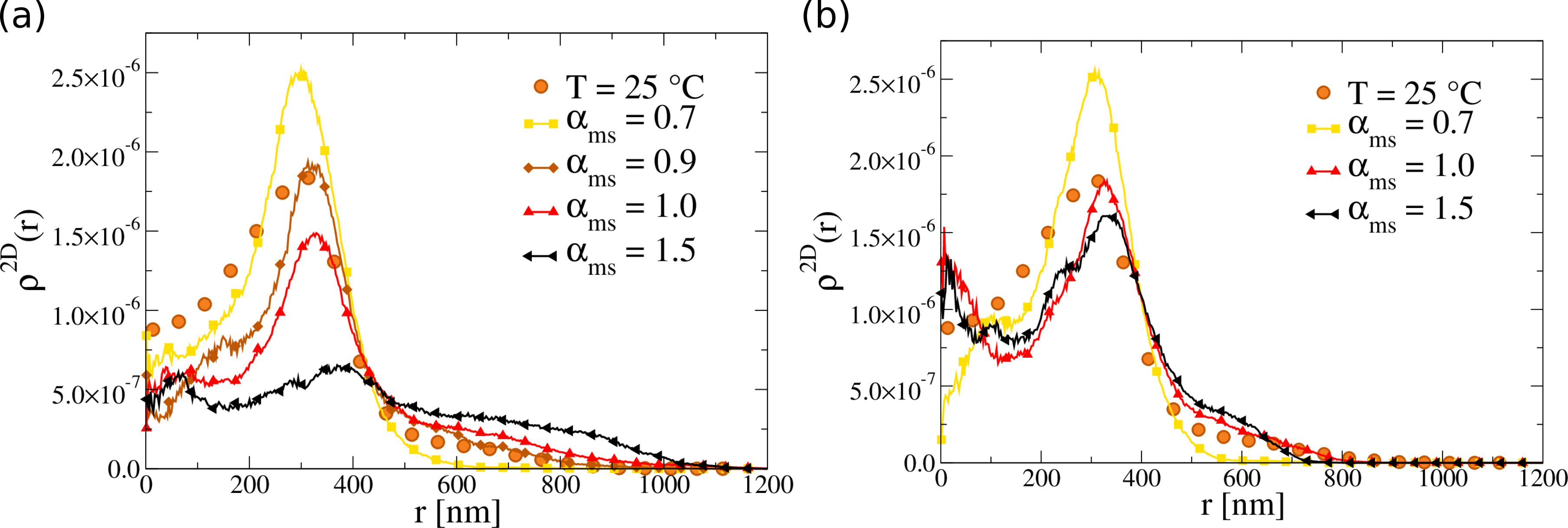}
  \caption{ Experimental 2D density profile at 25~\textdegree C (symbols) and numerical 2D density profiles (lines) calculated for an unbonded swollen microgel $\alpha_{\text{mm}}=0.0$ on surfaces at different $\alpha_{\text{ms}}$ values at (a) long times and (b) short times (in the window $5.5\times 10^{6} - 8\times 10^{6}$ steps). The profile with $\alpha_{\text{ms}}=0.9$ overall gives the best fit of the experimenal data.}
 \label{fig:mwa-selection}
\end{figure}

In order to find the optimal value of the monomer-surface attraction parameter $\alpha_{\text{ms}}$, which best approximates the hydrophobic HDMS treated surface in experiments, we performed simulations of unbonded microgels in a swollen state ($\alpha_{\text{mm}}=0.0$) near surfaces with different deegres of attraction $\alpha_{\text{ms}}$. Figure~\ref{fig:mwa-selection}(a) shows the experimental density profiles at 25~\textdegree C for the case of the hydrophobic surface compared with the calculated profiles for different $\alpha_{\text{ms}}$ values. The simulated density profile with $\alpha_{\text{ms}}=0.9$ is the one which more closely matches the experimental curve: it has a similar height of the peak and tail extension; still, some under-estimation at short distances is observed. As discussed in the main text, a better agreement is obtained by adding a fraction of bonded monomers to the surface, to mimic the experimental situation. 

Another important feature to take into account is that, with increasing  $\alpha_{\text{ms}}$,  equilibration of the system takes longer and longer. Fig.~\ref{fig:mwa-selection}(b) shows again the density profiles but now calculated at short times (from step $5.5\times 10^{6}$ to $8\times 10^{6}$ ) for different degrees of attraction.  At this point, we clearly see that the profile with the longest tail does not actually correspond to the most hydrophobic case. This is due to the fact that monomer-surface attached pairs take longer to break, expand and reform, so full equilibration takes a long time.

\FloatBarrier
\subsection{Effect of the monomer-surface attraction parameter on an anchored microgel}
\begin{figure}[ht]
\centering
  \includegraphics[width=0.6\textwidth]{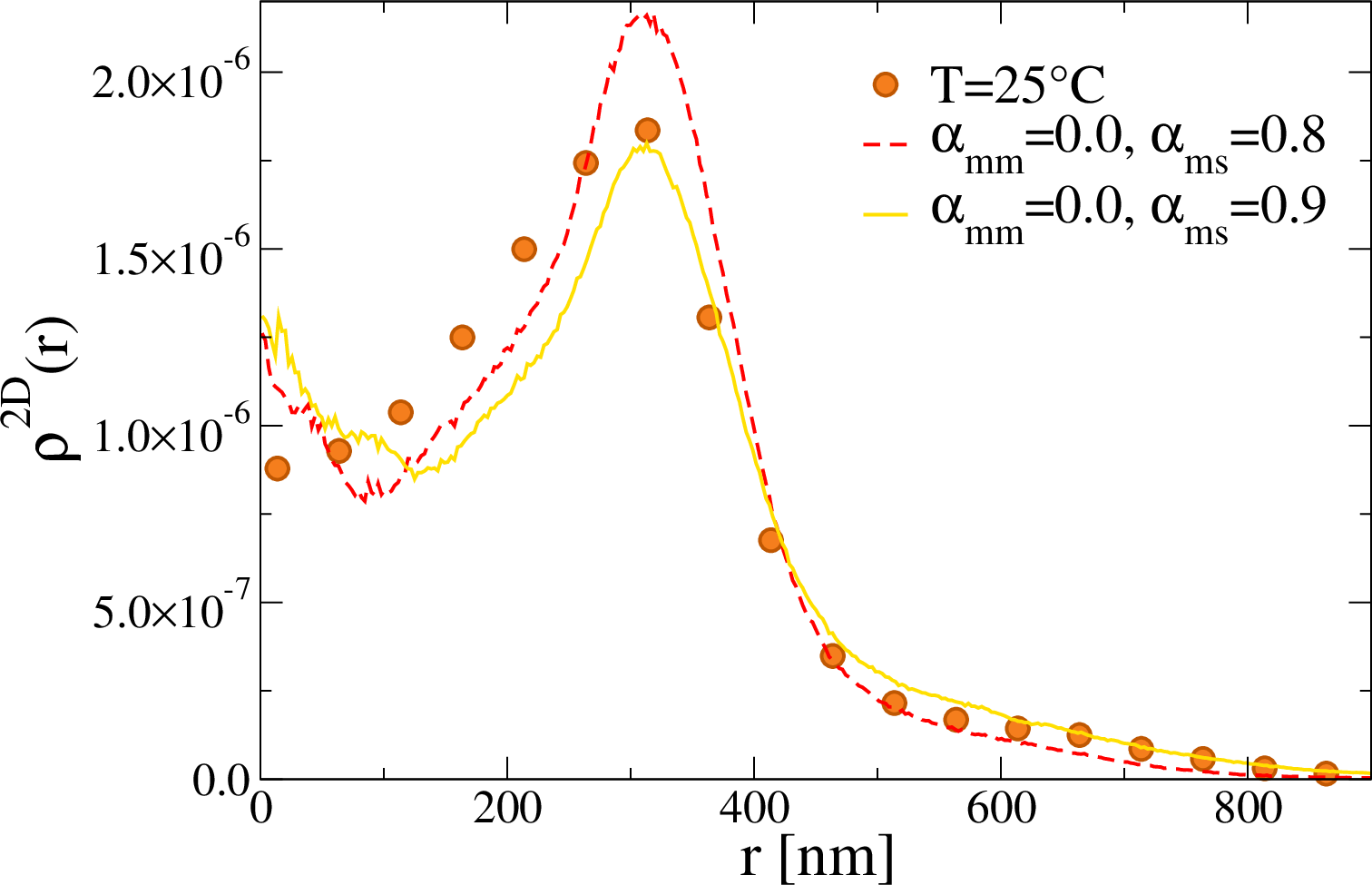}
  \caption{Experimental density profile T=25~\textdegree C (symbols)  and numerical density profiles (lines) of the bonded ($b=200$), swollen microgel  ($\alpha_{\text{ms}}=0.0$ ) close to an attractive surface with $\alpha_{\text{ms}}=0.8$ (dashed red line) and $\alpha_{\text{ms}}=0.9$ (solid yellow line).}
 \label{fig:mwa08-mwa09}
\end{figure}

Simulations of an unbonded microgel close to a hydrophobic surface suggested that the monomer-surface parameter should be close to $\alpha_{\text{ms}} \sim 0.9$. To verify that this value is correct also for a bonded microgel (with $b=200$ bonds, as the one used in the main text), we performed additional simulations both for $\alpha_{\text{ms}} = 0.9$ and $\alpha_{\text{ms}} =0.8$. Figure~\ref{fig:mwa08-mwa09} shows the 2D density profiles at both values of $\alpha_{\text{ms}}$ at T=25~\textdegree C. We find that below the VPT temperature, the data for $\alpha_{\text{ms}}=0.8$ overestimate the peak and underestimate the tail, differently from data for $\alpha_{\text{ms}}=0.9$, confirming that this is the optimal value to best capture the hydrophobicity of the experimental surface, even for anchored microgels.

\FloatBarrier
\subsection{Effect of the surface selection during the anchoring procedure on microgels above the VPTT for hydrophobic surfaces}

\begin{figure}[ht]
\centering
  \includegraphics[width=0.65\textwidth]{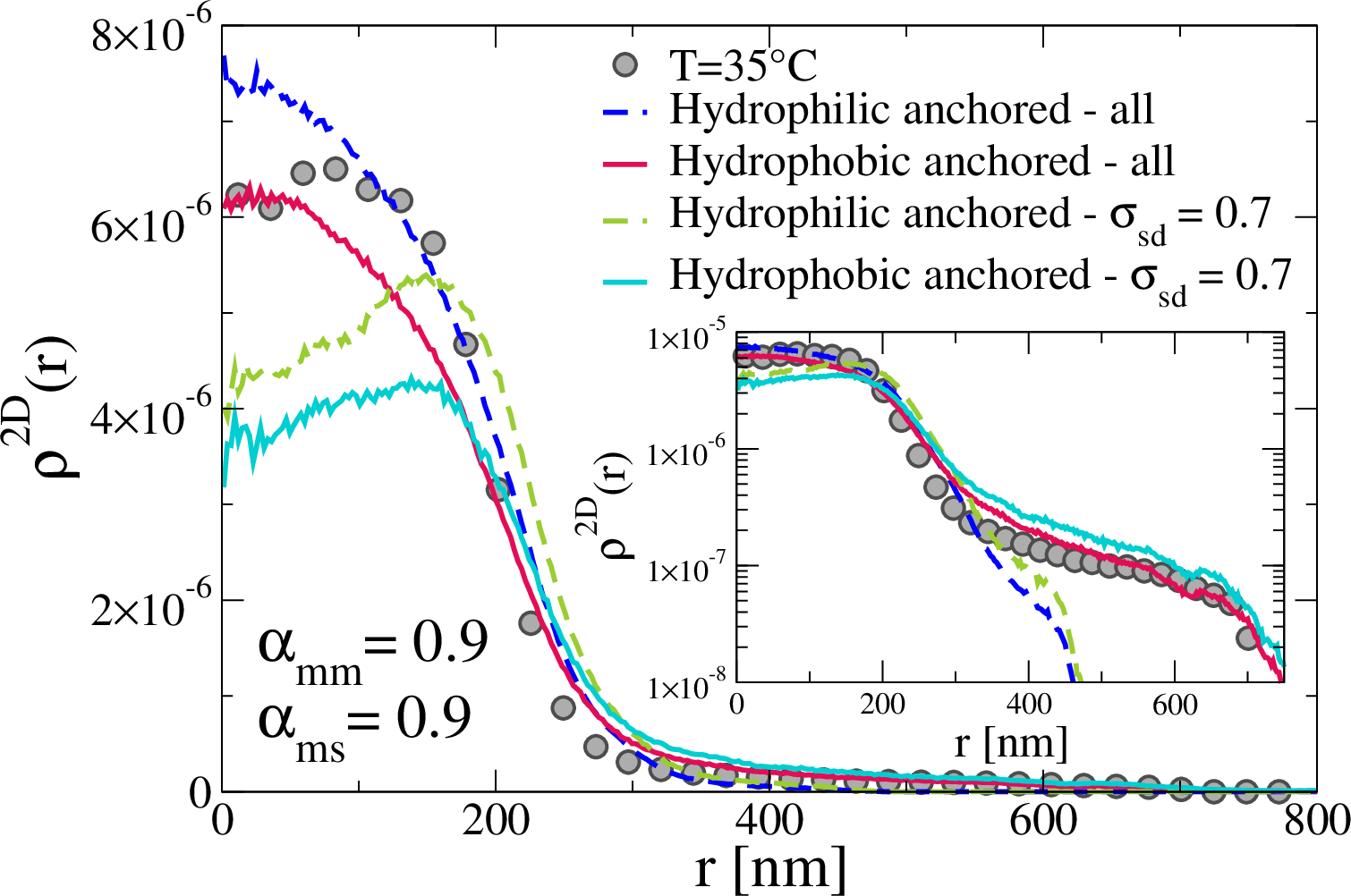}
  \caption{Experimental density profile (symbols) at 35~\textdegree C and numerical density profiles of a bonded microgel anchored in a hydrophilic surface (dashed line) and a hydrophobic surface (solid line) when seeing ``all'' monomers or a fraction of them following the distributions from Fig.~\ref{fig:hst3d_norm}. Inset: Logarithmic $y$-scale to show the density profiles long distance extension. }
 \label{fig:rebond}
\end{figure}

We look at the influence of the surface selection during the anchoring procedure on the microgel at high temperatures and close to a hydrophobic surface.
The anchoring procedure, explained in the main text, was followed using the main two surfaces $\alpha_{\text{ms}}=0.0, 0.9$; i.e. the microgel was pushed and anchored to surfaces either completely hydrophilic $\alpha_{\text{ms}}=0.0$ or considerably hydrophobic $\alpha_{\text{ms}}=0.9$; after bonding, the monomer-surface parameter was changed to $\alpha_{\text{ms}}=0.9$ for the first case.
During the anchoring on a hydrophobic surface, the microgel expands over it due to the attractive interaction $\alpha_{\text{ms}}=0.9$. This results in an increase of the number of suitable monomers to be anchored whose positions also get further away from the microgels center of mass. After anchoring and simulating above the VPT $\alpha_{\text{mm}}=0.9$, the anchored monomers far away from the microgels center of mass hamper the collapse hence extending the profile.
Figure~\ref{fig:rebond} shows the density profiles of the microgel anchored respectively to a hydrophilic and a hydrophobic surface. The profile tail extension is found to be significantly larger when anchoring is done from a hydrophobic surface. The increase of extension is also reflected in a decrease  of the profile at shorter distances.  
Additionally, we show the comparison of the profiles when seen the full microgel or a fraction of its monomers as in the hydrophilic scenario ($\sigma_{\text{sd}}=0.7$). As mentioned in the main text, the peak disappears and the amplitude at short distances increase in the full microgel profiles, resembling closer the experiments.
Furthermore, the fitting of the tail extension for the fully seen microgel anchored on a hydrophobic surface captures better the experimental profiles respect the partially labeled one.

\bibliography{ref.bib}